\pdfoutput=1

\documentclass[11pt,twoside,a4paper,cmspaper,final,collab]{cms-tdr}
\def\svnVersion{484858P}\def\cmsCernNoTag{CERN-EP-2017-343}\def\cmsCernDate{\today}\def\cmsMessage{Published in Physics Letters B as \href{http://dx.doi.org/10.1016/j.physletb.2018.10.056}{\doi{10.1016/j.physletb.2018.10.056.}}}

\begin{document}\cmsNoteHeader{HIG-17-008}

\hyphenation{had-ron-i-za-tion}
\hyphenation{cal-or-i-me-ter}
\hyphenation{de-vices}
\RCS$HeadURL: svn+ssh://svn.cern.ch/reps/tdr2/papers/HIG-17-008/trunk/HIG-17-008.tex $
\RCS$Id: HIG-17-008.tex 477190 2018-10-05 19:18:54Z alverson $
\newlength\cmsFigWidth
\ifthenelse{\boolean{cms@external}}{\setlength\cmsFigWidth{0.98\columnwidth}}{\setlength\cmsFigWidth{0.7\textwidth}}
\ifthenelse{\boolean{cms@external}}{\providecommand{\cmsLeft}{top\xspace}}{\providecommand{\cmsLeft}{left\xspace}}
\ifthenelse{\boolean{cms@external}}{\providecommand{\cmsRight}{bottom\xspace}}{\providecommand{\cmsRight}{right\xspace}}
\newlength\cmsTabSkip\setlength{\cmsTabSkip}{1ex}
\ifthenelse{\boolean{cms@external}}{\providecommand{\cmsResize}[1]{#1}}{\providecommand{\cmsResize}[1]{\resizebox{\textwidth}{!}{#1}}}

\newcommand{\HH}{\ensuremath{{\PH\PH}}\xspace}
\newcommand{\matcH}{H\xspace}
\newcommand{\ppHHbbgg}{\ensuremath{\Pp\Pp\to\PH\PH\to\gamma\gamma\bbbar}\xspace}
\newcommand{\ppXHHbbgg}{\ensuremath{\Pp\Pp\to \text{X}\to\PH\PH\to\gamma\gamma\bbbar}\xspace}
\newcommand{\HHbbgg}{\ensuremath{\PH\PH\to\gamma\gamma\bbbar}\xspace}
\newcommand{\mHH}{\ensuremath{m_{\PH\PH}}\xspace}
\newcommand{\Hgg}{\ensuremath{\PH\to\gamma\gamma}\xspace}
\newcommand{\Hbb}{\ensuremath{\PH\to \bbbar}\xspace}

\newcommand{\ggHH}{\ensuremath{\Pg\Pg\PH\PH}\xspace}

\newcommand{\ggH}{\ensuremath{\Pg\Pg\PH}\xspace}
\newcommand{\ttH}{\ensuremath{\ttbar\PH}\xspace}
\newcommand{\bbH}{\ensuremath{\bbbar\PH}\xspace}
\newcommand{\VH}{\text{V}\ensuremath{\PH}\xspace}
\newcommand{\VBFH}{\text{VBF}~\PH\xspace}

\newcommand{\sigmaHH}{\ensuremath{\sigma_{\Pg\Pg \to \PH\PH}}\xspace }
\newcommand{\sigmaHHSM}{\ensuremath{\sigma_{\Pg\Pg \to \PH\PH}^\text{SM}\xspace}}
\newcommand{\sigmaHHBSM}{\ensuremath{\sigma_{\Pg\Pg \to \PH\PH}^\text{BSM}\xspace}}
\newcommand{\sigmaHHSMVBF}{\ensuremath{\sigma_{\text{VBF}~\PH\PH}^\text{SM}\xspace}}
\newcommand{\sigmaHHBSMVBF}{\ensuremath{\sigma_{\text{VBF}~\PH\PH}^\text{BSM}\xspace}}
\newcommand{\muHH}{\ensuremath{\mu_{\HH}}\xspace}
\newcommand{\muHHext}{\ensuremath{\mu_{\HH}^\text{ext}\xspace}}

\newcommand{\bbgg}{\ensuremath{\gamma\gamma\bbbar}\xspace}
\newcommand{\QCD}{\ensuremath{n\gamma + \text{jets}}\xspace}
\newcommand{\mt}{\ensuremath{m_\cPqt}\xspace}
\newcommand{\mx}{\ensuremath{m_\text{X}\xspace}}
\newcommand{\Mtilde}{\ensuremath{\widetilde{M}_{\mathrm{X}}}\xspace}
\newcommand{\koM}{\ensuremath{\kappa/\overline{\Mpl}}\xspace}
\newcommand{\Mgg}{\ensuremath{m_{\gamma\gamma}}\xspace}
\newcommand{\Mjj}{\ensuremath{m_\text{jj}\xspace}}
\newcommand{\thetastar}{\ensuremath{\theta^\text{CS}_{\HH}}}
\newcommand{\acosthetastar}{\ensuremath{\abs{\cos {\theta^\text{CS}_{\HH}}}\xspace}}
\newcommand{\acosthetabb}{\ensuremath{\abs{\cos {\theta_\text{jj}}}}\xspace}
\newcommand{\acosthetagg}{\ensuremath{\abs{\cos {\theta_{\gamma \gamma}}}\xspace}}
\newcommand{\Mggjj}{\ensuremath{m_{\gamma\gamma \text{jj}}}}
\newcommand{\ptgg}{\ensuremath{\pt^{\gamma\gamma}}}
\newcommand{\ptjj}{\ensuremath{\pt^{\text{jj}}}}

\newcommand{\AxE}{\ensuremath{\text{A} \, \epsilon}\xspace}

\newcommand{\kapt}{\ensuremath{\kappa_{\PQt}}\xspace}
\newcommand{\kapl}{\ensuremath{\kappa_{\lambda}}\xspace}
\newcommand{\ctwo}{\ensuremath{c_2}\xspace}
\newcommand{\cg}{\ensuremath{c_{\Pg}}\xspace}
\newcommand{\cgg}{\ensuremath{c_{2\Pg}}\xspace}
\newcommand{\LambdaR}{\ensuremath{\Lambda_\mathrm{R}}\xspace}
\newcommand{\ptgone}{\ensuremath{\pt^{\gamma 1}}\xspace}
\newcommand{\ptgtwo}{\ensuremath{\pt^{\gamma 2}}\xspace}
\newcommand{\mH}{\ensuremath{m_{\PH}}\xspace}
\newcommand{\lbdSM}{\ensuremath{\lambda^\mathrm{SM}_{\PH\PH\PH}}\xspace}
\newcommand{\lbdHHH}{\ensuremath{\lambda_{\PH\PH\PH}}\xspace}
\newcommand{\Rj}{\ensuremath{R_\text{j}}}
\newcommand{\DRgj}{\ensuremath{\Delta R_{\gamma \mathrm{j}}}\xspace}
\newcommand{\Dphigj}{\ensuremath{\Delta \phi_{\gamma \mathrm{j}}}\xspace}
\newcommand{\Detagj}{\ensuremath{\Delta \eta_{\gamma \mathrm{j}}}\xspace}
\newcommand{\yt}{\ensuremath{y_{\PQt}}\xspace}
\newcommand{\ytSM}{\ensuremath{y_{\PQt}^\text{SM}\xspace}}

\newcommand\plotsize{0.40}

\cmsNoteHeader{HIG-17-008}

\title{Search for Higgs boson pair production in the $\gamma\gamma\bbbar$ final state in $\Pp\Pp$ collisions at $\sqrt{s}=13\TeV$}

\date{\today}

\abstract{
A search is presented for the production of a pair of Higgs bosons, where one decays into two photons and the other one into a bottom quark-antiquark pair. The analysis is performed using proton-proton collision data at $\sqrt{s} = 13\TeV$ recorded in 2016 by the CMS detector at the LHC, corresponding to an integrated luminosity of 35.9\fbinv. The results are in agreement with standard model (SM) predictions.
In a search for resonant production, upper limits are set on the cross section for new spin-0 or spin-2 particles.
For the SM-like nonresonant production hypothesis, the data exclude a product of cross section and branching fraction larger than 2.0\unit{fb} at 95\% confidence level (\CL), corresponding to about 24 times the SM prediction. Values of the effective Higgs boson self-coupling $\kappa_\lambda$ are constrained to be within the range $-11 < \kappa_\lambda < 17$ at 95\% \CL, assuming all other Higgs boson couplings are at their SM value. The constraints on $\kappa_\lambda$ are the most restrictive to date.
}

\hypersetup{%
pdfauthor={CMS Collaboration},%
pdftitle={Search for Higgs boson pair production in the ggbb final state in pp collisions at sqrt(s) = 13 TeV},%
pdfsubject={CMS},%
pdfkeywords={CMS, physics, Higgs, photons, b-jets}}

\maketitle

\section{Introduction}\label{sec:intro}

The discovery of a particle with a mass of about $125\GeV$, with properties compatible with those expected for the
Higgs (\PH) boson of the standard model (SM)~\cite{HiggsdiscoveryAtlas, Chatrchyan:2012ufa, Chatrchyan:2013lba}, has stimulated
interest in the detailed exploration and understanding of the origin of the Brout--Englert--Higgs (BEH) mechanism \cite{Higgs1, Higgs:1964pj}.
The production of a pair of Higgs bosons (\HH) is a rare process that is sensitive to the structure of the BEH potential through the Higgs boson's self-coupling mechanism.
In the SM, the corresponding production cross section via gluon-gluon fusion in proton-proton ($\Pp\Pp$) collisions at $\sqrt{s} = 13\TeV$ is predicted to be $\sigmaHH = 33.5^{+2.5}_{-2.8}\unit{fb}$~\cite{deFlorian:2016spz, deFlorian:2013jea,Baglio:2012np}, a value beyond the reach of the analyses based on the current integrated luminosity of the CERN LHC program.

Many theories beyond the SM (BSM) suggest the existence of heavy particles that can couple to a pair of Higgs bosons. These particles could appear as a resonant contribution to the invariant mass of the \HH system and induce a significant increase of the $\HH$ production cross section with respect to the SM.
For example, models with warped extra dimensions
(WED)~\cite{Randall:1999ee} postulate the existence of compactified extra spatial dimensions. They predict new resonances decaying to a Higgs boson pair, including spin-0 radions $R$~\cite{Goldberger:1999uk}
and spin-2 Kaluza--Klein gravitons (KK graviton)~\cite{Davoudiasl:1999jd}.
The benchmark scenario considered in this paper, the bulk Randall--Sundrum (RS) model~\cite{Fitzpatrick:2007qr}, assumes that all fields can propagate in the extra dimension.
Models with an extended Higgs sector also predict a spin-0 resonance that decays to a pair of SM Higgs bosons, if sufficiently massive.
Examples of such models are the singlet extension of the SM~\cite{O'Connell:2006wi}, the two-Higgs-doublet models~\cite{Branco:2011iw} (in particular, the minimal supersymmetric model~\cite{Djouadi:2005gj}), and the Georgi--Machacek model~\cite{Georgi:1985nv}. Many of these models predict that heavy scalar production occurs predominantly through the gluon-gluon fusion process. The Lorentz structure of the effective coupling between the scalar and the gluon is the same for a radion or a heavy Higgs boson. Therefore, the kinematics for the production of a radion or an additional Higgs boson are essentially the same, provided the spin-0 resonance is narrow. The radion results can therefore be applied to constrain this class of models.

If the new particles are too heavy to be observed through a direct search, they may contribute to \HH production through virtual processes and lead to an enhancement of the cross section with respect to the SM prediction (as discussed, \eg, in Ref.~\cite{Dawson:2015oha}). In addition, different BSM models can modify the Higgs boson's fundamental couplings and impact \HH production in gluon-gluon fusion~\cite{Grober:2010yv} (\ggHH) and vector boson fusion (VBF)~\cite{deFlorian:2016spz, Baglio:2012np}.

This letter describes a search for the production of pairs of Higgs bosons via $\ppHHbbgg$
using a data sample of 35.9\fbinv collected by the CMS experiment in 2016.
Both nonresonant and resonant production are explored, with the search for a narrow resonance X conducted at masses $\mx$ between 260 and 900\GeV.
The fully reconstructed $\bbgg$ final state
combines the large SM branching fraction ($\mathcal{B}$) of the $\Hbb$ decay
with the comparatively low background and good mass resolution of the $\PH \to \gamma \gamma$ channel,
yielding a total $\mathcal{B}(\HHbbgg)$ of 0.26\%~\cite{deFlorian:2016spz}.
The search uses the mass spectra of the diphoton ($\Mgg$), dijet ($\Mjj$),
and four-body systems ($\Mggjj$), as well as the associated helicity angles, to provide discrimination
between the $\HH$ production signal and the other SM processes. The \ggHH production process is studied in detail and the sensitivity of CMS data to the nonresonant VBF production mechanism is investigated for the first time.

Searches in the same and complementary final states such as $\HH \to \bbbar\bbbar$ or $\HH \to \tau^{+}\tau^{-}\bbbar$ were performed in the past by the ATLAS \cite{Aad:2014yja, Aad:2015uka, Aad:2015xja, Aaboud:2016xco} and CMS \cite{Khachatryan:2016sey, Khachatryan:2015year, Sirunyan:2017isc,  Sirunyan:2017tqo, Sirunyan:2017djm, Sirunyan:2017guj} Collaborations at $\sqrt{s}=8$ and 13\TeV.

\section{The CMS detector} \label{sec-detector}

The CMS detector, its coordinate system, and the main kinematic variables used in the analysis are described in detail
in Ref.~\cite{Chatrchyan:2008zzk}.
The central feature of the CMS apparatus is a superconducting solenoid, of 6\unit{m} internal diameter, providing a
magnetic field of 3.8\unit{T}. A silicon pixel and strip tracker covering the pseudorapidity range $\abs{\eta}< 2.5$, an
electromagnetic calorimeter (ECAL) made of lead tungstate crystals, and a brass and scintillator hadron calorimeter reside within the field volume.
Forward calorimeters extend the pseudorapidity coverage above $\abs{\eta} = 3.0$.
Muons are detected in gas-ionization chambers embedded in the steel flux-return yoke outside the solenoid.
The first level of the CMS trigger system, composed of special hardware processors, uses information from the calorimeters and muon detectors to select
the most interesting events in a time interval of less than 4\mus. The high-level trigger further decreases the event rate, from around 100\unit{kHz}
to less than 1\unit{kHz}, before data storage \cite{Khachatryan:2016bia}.

\section{Simulated events} \label{sec:samples}

Signal samples are simulated at leading order (LO) using \MGvATNLO 2.3.2 \cite{Alwall:2014hca, Hespel:2014sla, Frederix:2014hta} interfaced with {LHAPDF6} \cite{Buckley:2014ana}. The next-to-leading-order (NLO) parton distribution function (PDF) set
{PDF4LHC15\_NLO\_MC} is used~\cite{Carrazza:2015hva, Butterworth:2015oua, Dulat:2015mca, Harland-Lang:2014zoa, Ball:2014uwa}.
The models describe the production through gluon-gluon fusion of particles with narrow width (set to 1\MeV) that decay to two Higgs bosons with the mass of $m_{\PH} = 125\GeV$ \cite{Aad:2015zhl}. Events are generated either for spin-0 radion production, or spin-2 KK graviton production, as predicted by the bulk RS model. For each spin hypothesis 16 mass points are generated within the range $260 \leq \mx \leq 900\GeV$ in steps of 10\GeV for $\mx$ between 260 and 300\GeV, and in steps of 50\GeV for $\mx$ above 300\GeV.

\begin{figure*}[hbt]
\centering
\includegraphics[scale=0.33]{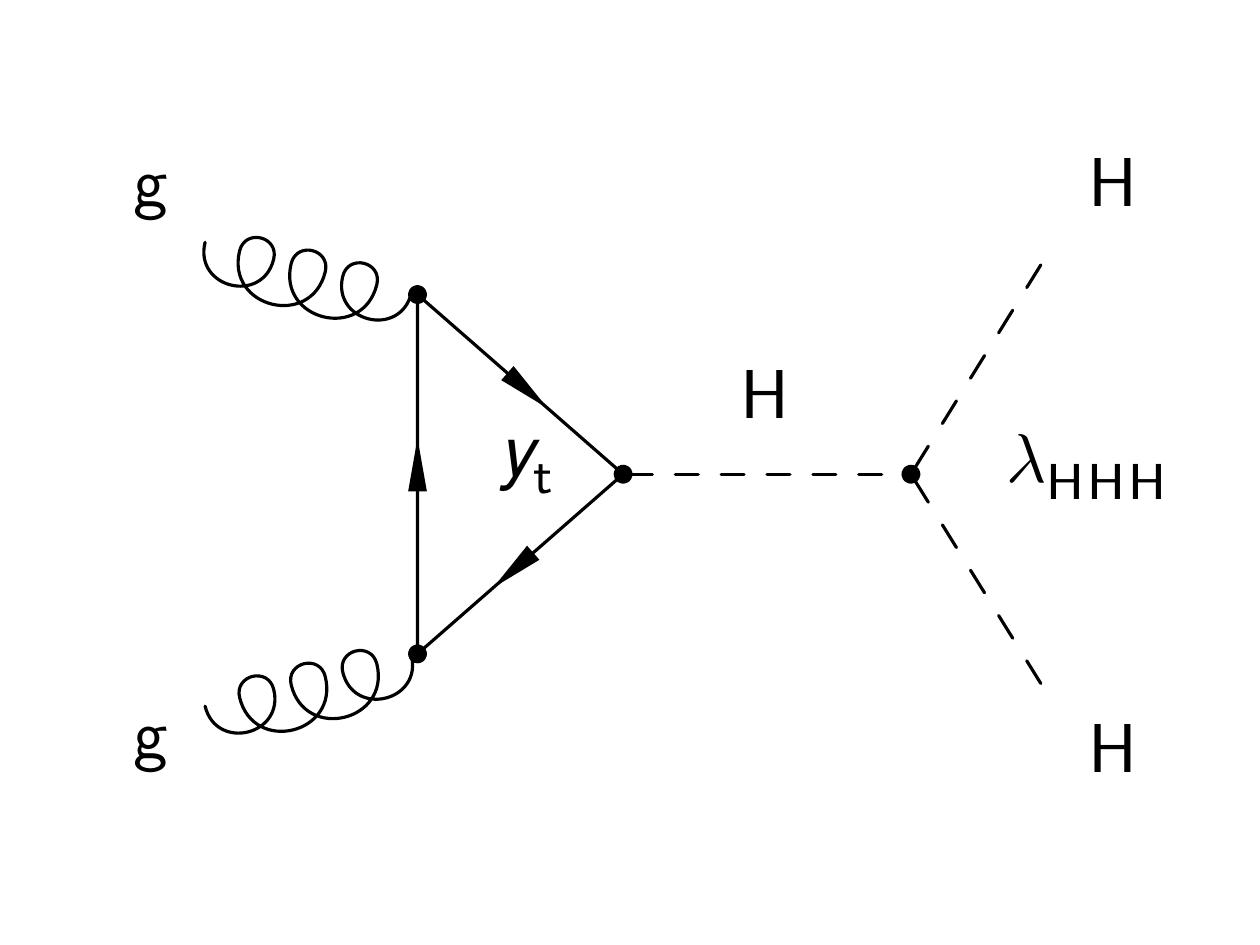}
\includegraphics[scale=0.33]{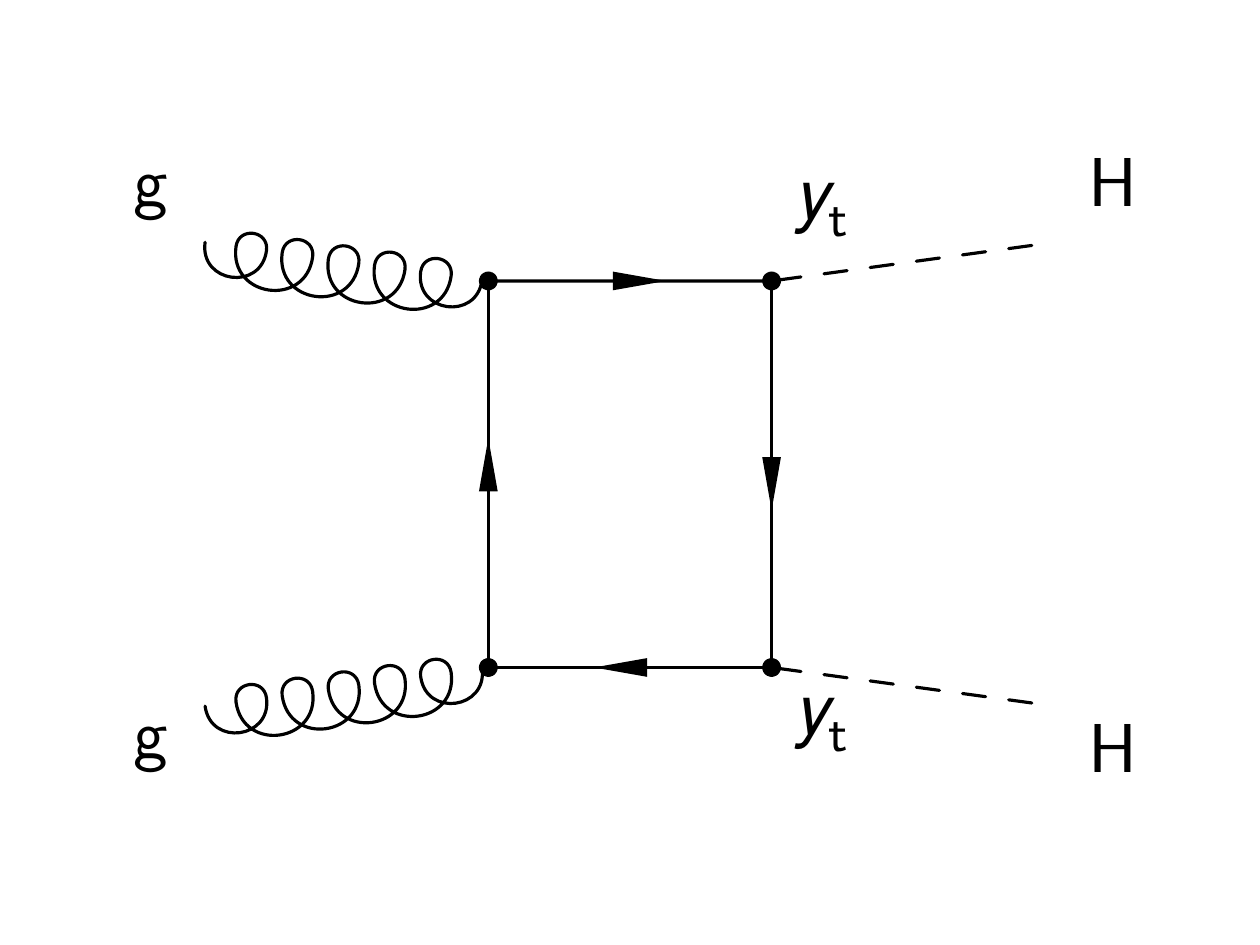}\\
\includegraphics[scale=0.33]{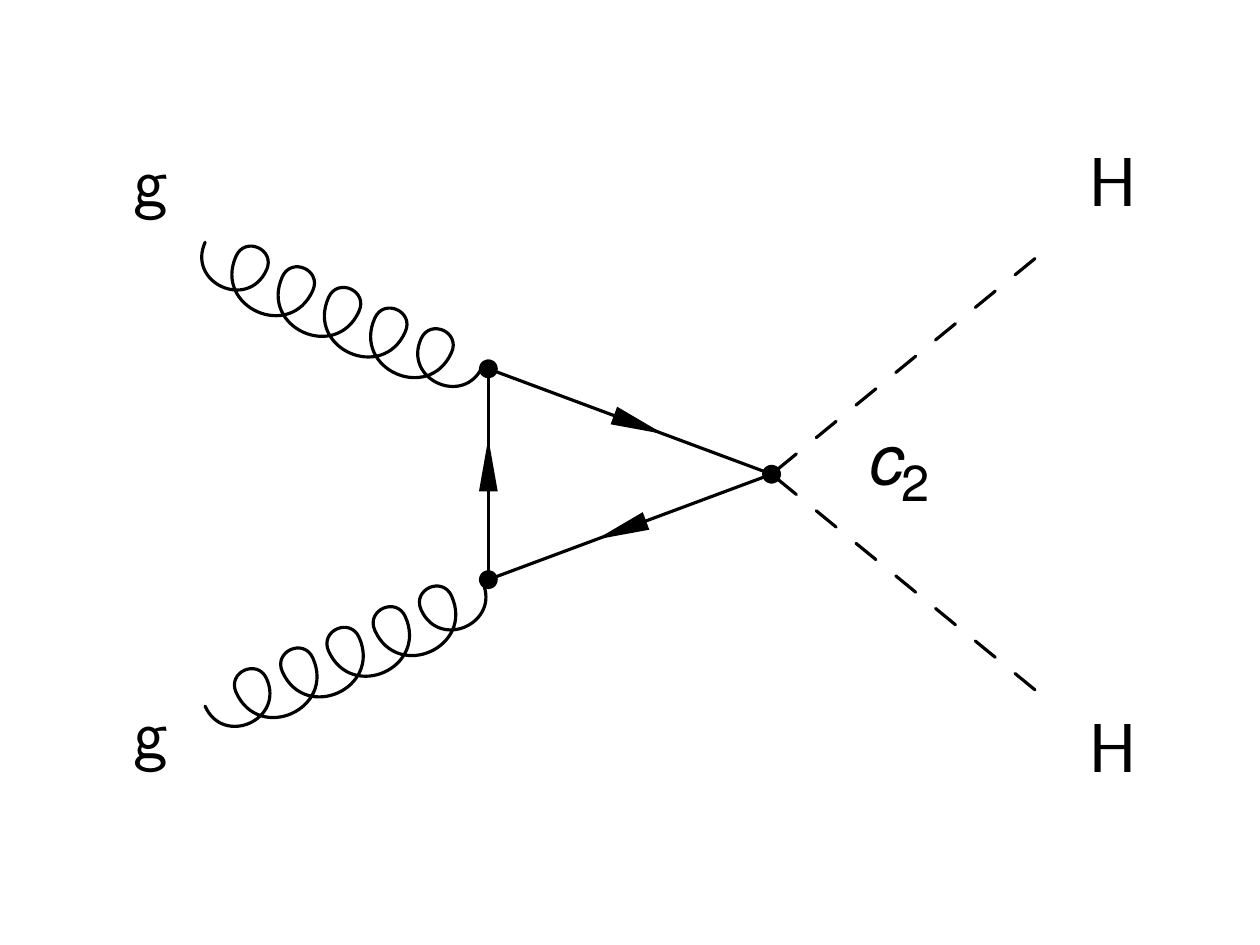}
\includegraphics[scale=0.33]{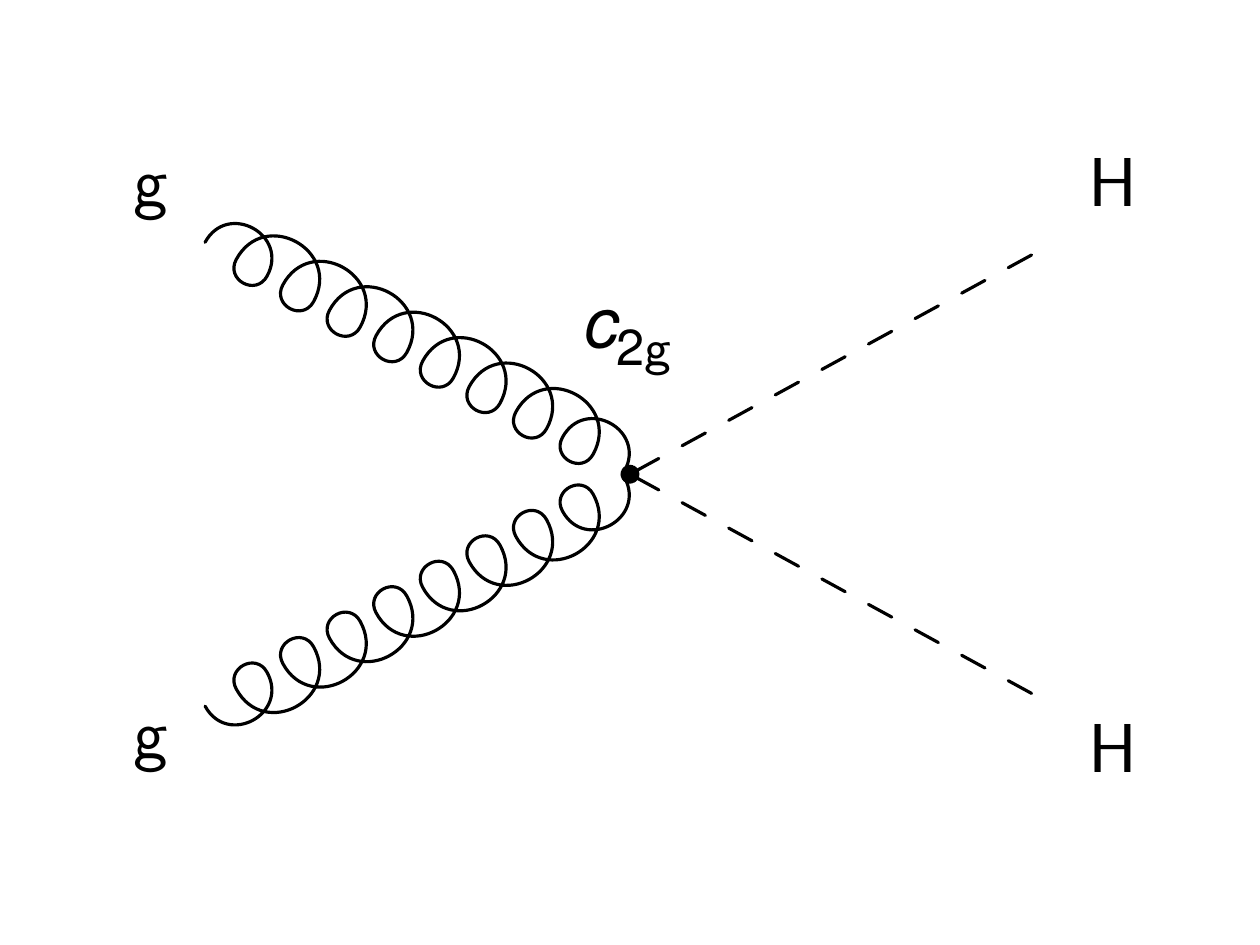}
\includegraphics[scale=0.33]{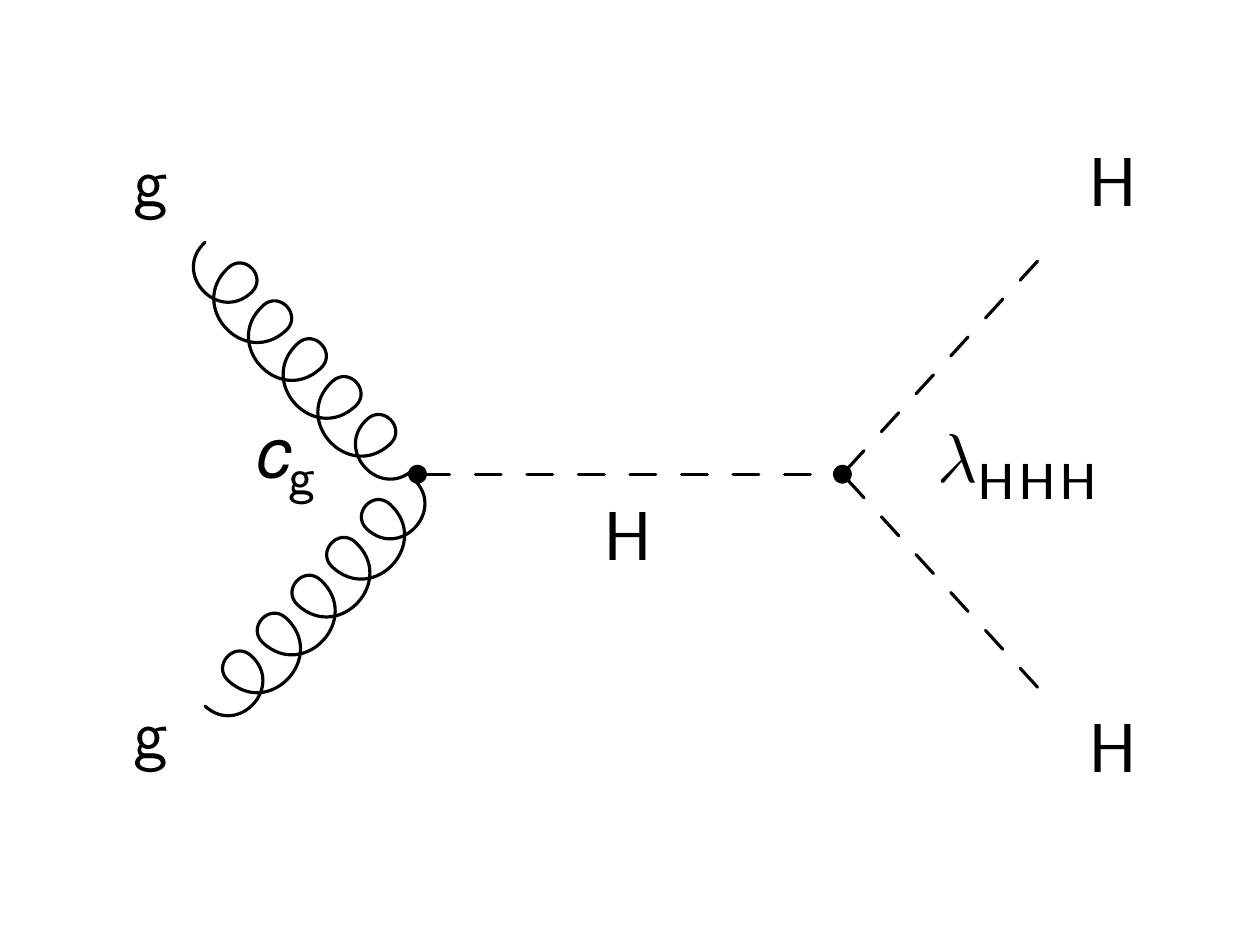}\\
\caption{Feynman diagrams that contribute to \ggHH at LO. Top diagrams correspond to SM-like processes, referred to as box and triangle diagrams, respectively. The bottom diagrams correspond to pure BSM processes: the first exploits the contact interaction of two Higgs bosons with two top quarks and the last two describe contact interactions between the \PH boson and gluons.\label{fig:dia}}
\end{figure*}

In the nonresonant case we use the effective field theory approach and notations from Refs.~\cite{deFlorian:2016spz, Carvalho:2016rys}. First we consider two SM coupling modifiers: $\kapl \equiv \lbdHHH/\lbdSM$, which measures deviations of the Higgs boson trilinear coupling $\lbdHHH$ from its SM expectation, $\lbdSM \equiv \mH^2 /(2 v^2) = 0.129$; and $\kapt \equiv \yt/\ytSM$, which measures deviations of the top quark Yukawa coupling $\yt$ from its SM expectation $\ytSM = \sqrt{2}\,\mt/v \approx 1.0$. Here $v = 246\GeV$ denotes the vacuum expectation value of the Higgs boson, $\mH$ its mass, and $\mt$ denotes the top quark mass.
Second, we also consider couplings not found in the SM that are derived from dimension-6 operators: contact interactions between two Higgs bosons and two top quarks ($\ctwo$), between one Higgs boson and two gluons ($\cg$), and between two Higgs bosons and two gluons ($\cgg$). We define these three parameters in such a way that their values are zero within the SM.
The most general production is described by a modification of the SM Lagrangian, the relevant part of which, labelled as $\mathcal{L}_{\HH}$, is given in the following equation~\cite{Giudice:2007fh}:
\ifthenelse{\boolean{cms@external}}{
\begin{multline}
\mathcal{L}_{\HH} =  {\kapl}\,  \lbdSM v\, \matcH^3
  - \frac{ \mt }{v}\bigl({\kapt} \,   \matcH  +  \frac{\ctwo}{v}   \, \matcH^2 \bigr) \,\bigl( \overline{\PQt}_\text{L}\PQt_\text{R} + \text{h.c.}\bigr)  \\+ \frac{1}{4} \frac{\alpS}{3 \pi v} \bigl(   \cg \, \matcH -  \frac{\cgg}{2 v} \, \matcH^2  \bigr) \,  G^{\mu \nu}G_{\mu\nu},
\label{eq:lag}
\end{multline}
}{
\begin{multline}
\mathcal{L}_{\HH} =  {\kapl}\,  \lbdSM v\, \matcH^3
  - \frac{ \mt }{v}\bigl({\kapt} \,   \matcH  +  \frac{\ctwo}{v}   \, \matcH^2 \bigr) \,\bigl( \overline{\PQt}_\text{L}\PQt_\text{R} + \text{h.c.}\bigr)  + \frac{1}{4} \frac{\alpS}{3 \pi v} \bigl(   \cg \, \matcH -  \frac{\cgg}{2 v} \, \matcH^2  \bigr) \,  G^{\mu \nu}G_{\mu\nu},
\label{eq:lag}
\end{multline}
}
where  $\PQt_\text{L}$ and $\PQt_\text{R}$ are the top quark fields with left and right chiralities, respectively. The $\matcH$ denotes the physical Higgs boson field, $G^{\mu \nu}$ is the gluon field strength tensor, and $\alpS$ is the strong coupling constant. The notation $h.c.$ is used for the Hermitian conjugate. Five main Feynman diagrams, shown in Fig. \ref{fig:dia}, contribute to \ggHH at LO.

The cross section is expressed at LO as a function of five BSM parameters ($\kapl$, $\kapt$, $\ctwo$, $\cg$, and $\cgg$) \cite{deFlorian:2016spz, Carvalho:2016rys} and this parametrization is approximately extended to the next-to-next-to-leading order matched to the next-to-next-to-leading log in quantum chromodynamics (QCD) using a global k-factor \cite{deFlorian:2013jea, Grigo:2014jma, deFlorian:2016spz} that has uncertainties coming from the PDF, missing orders, $\alpS$ and finite top quark mass effects. There is a dependence of the k-factors on $\mHH$ related to the finite top quark mass effects \cite{Borowka:2016ypz}. Within the region of sensitivity of this analysis, $\mHH < 900\GeV$, any effect is covered by the total k-factor uncertainty. The departure of the parameters from their SM values can change the total cross section by a few orders of magnitude. Furthermore, the kinematic properties of the \HH final state are modified, which affects the final sensitivity through the modification of the acceptance and efficiency of the experimental analysis.

To avoid a prohibitively large number of samples to simulate and to analyze, we use the method proposed in Refs.~\cite{Dall'Osso:2015aia, Carvalho:2016rys, deFlorian:2016spz} to partition the parameter space into 12 regions with distinct kinematics, referred to as clusters.
Each of the clusters with its \HH kinematics can be represented by a point in the 5D parameter space that is referred to as a \textit{benchmark hypothesis}. We simulate the 12 benchmark hypotheses together with two additional ones -- one assuming all parameters to be SM ones (referred to as \textit{SM benchmark hypothesis}) and the other using identical assumptions except for a vanishing Higgs boson self-coupling (referred to as $\kapl = 0$). The list of benchmark hypotheses is provided in Table \ref{tab:bench}.
An additional \HH sample is produced via the VBF mechanism using SM couplings.

The ensemble of events obtained by combining all 14 gluon-gluon initiated samples covers the possible kinematic configurations of the effective field theory parameter space. These events can therefore subsequently be reweighted using the procedure derived in Ref.~\cite{Carvalho:2017vnu} to model any desired point in the full 5D parameter space. In this procedure, an event-by-event weight is analytically calculated from the generator-level information on the \HH system.

\begin{table*}[h]
  \centering
\topcaption{Parameter values of nonresonant BSM benchmark hypotheses. The first two columns correspond to the SM and $\kapl = 0$ samples, respectively, while the next 12 correspond to the benchmark hypotheses identified using the method from Ref.~\cite{Dall'Osso:2015aia}. \label{tab:bench}}
\newcolumntype{R}{>{$}{r}<{$}}
\cmsResize{
  \begin{tabular}{l  R R R   R R R   R R R  R R R  R R}
\hline
&     \text{SM} & \kapl = 0 & 1 &   2 &    3 &    4 &   5 &   6 &   7 &    8 &   9 &   10 &  11 &   12 \\
\hline
$\kapl$  & 1.0 & 0.0  &  7.5 &  1.0 &  1.0 & -3.5 &  1.0 &  2.4 &  5.0 & 15.0 &  1.0 & 10.0 &  2.4 & 15.0 \\
$\kapt$  & 1.0 & 1.0  &  1.0 &  1.0 &  1.0 &  1.5 &  1.0 &  1.0 &  1.0 &  1.0 &  1.0 &  1.5 &  1.0 &  1.0 \\
$\ctwo$  & 0.0 & 0.0  & -1.0 &  0.5 & -1.5 & -3.0 &  0.0 &  0.0 &  0.0 &  0.0 &  1.0 & -1.0 &  0.0 &  1.0 \\
$\cg$    & 0.0 & 0.0  &  0.0 & -0.8 &  0.0 &  0.0 &  0.8 &  0.2 &  0.2 & -1.0 & -0.6 &  0.0 &  1.0 &  0.0 \\
$\cgg$   & 0.0 & 0.0  &  0.0 &  0.6 & -0.8 &  0.0 & -1.0 & -0.2 & -0.2 &  1.0 &  0.6 &  0.0 & -1.0 &  0.0 \\
\hline
\end{tabular}
}
\end{table*}

The dominant backgrounds to the $\bbgg$ final state are those in which two objects identified as photons (either prompt photons or jets misidentified as photons) are produced in association with jets (referred to as $\QCD$). The simulation of these final states is challenging due to large effects from higher orders in QCD~\cite{Faeh:2017fpp} and limited knowledge of fragmentation effects in the case of a jet misidentified as a photon.
In this analysis, these contributions are modeled entirely from data.

Single Higgs boson production in the SM, with two additional jets and with a subsequent decay of the Higgs boson to two photons, is also considered. In some cases, additional jets can be effectively initiated by \PQb quarks, but in others they can be initiated by lighter quarks and misidentified as a \PQb jet.
The considered processes --- gluon-gluon fusion (\ggH), VBF, and associated production with $\ttbar$ (ttH), $\bbbar$ (\bbH), and vector bosons (\VH) --- are sources of background in this analysis.
They are simulated using \MGvATNLO 2.2.2 for \VH and 2.3.3 for \bbH, and \POWHEG 2.0~\cite{Nason:2004rx, POWHEG_Frixione:2007vw, Alioli:2010xd, Bagnaschi:2011tu} at NLO for \ggH, \VBFH, and \ttH. All single-Higgs background samples are normalized to the SM cross section as recommended in Ref.~\cite{deFlorian:2016spz}.

All generated events are processed with \PYTHIA~8.212 \cite{Sjostrand:2014zea} with the tune CUETP8M1 \cite{Khachatryan:2015pea} for showering, hadronization, and the underlying event description, and \GEANTfour~\cite{Agostinelli:2002hh} for the simulation of the CMS detector response. The simulated events include multiple overlapping $\Pp\Pp$ interactions occurring in the same bunch crossing (pileup) as observed in the data.

\section{Data set and event selection}\label{sec:objects}

Events are selected using double-photon triggers, which require two photons with transverse momenta $\ptgone > 30\GeV$ and $\ptgtwo > 18\GeV$ for the leading and subleading photons, respectively. In addition, calorimeter-based isolation and shower shape requirements are imposed online on the two photons. Finally, the diphoton invariant mass is required to exceed 90\GeV.

In the offline selection, events are required to have at least one well-identified $\Pp\Pp$ collision vertex with a
position less than 24\cm away from the nominal interaction point in the $z$-direction. The primary vertex is identified by a multivariate analysis that was trained for the measurement of $\Hgg$ production \cite{Khachatryan:2014ira}. This analysis uses the momenta of the charged particle tracks associated with the vertex, and variables that quantify the vector and scalar balance of $\pt$ between the diphoton system
and the charged particle tracks associated with the vertex. The presence of at least two jets in the final state of this analysis
helps to correctly identify the primary vertex in more than 99.9\% of the simulated signal events.

\subsection{The \texorpdfstring{$\PH\to \gamma \gamma$}{Higgs to two photons} candidate}

Photons are identified using a multivariate technique that includes as inputs the $\pt$ of the electromagnetic shower, its longitudinal leakage into the hadron calorimeter, and its isolation from jet activity in the event. It was designed during the data taking at $\sqrt{s} = 8\TeV$ \cite{Khachatryan:2014ira, Khachatryan:2015iwa} and retrained with $\sqrt{s} = 13\TeV$ data. Identified photon candidates with a track matched to the ECAL cluster are rejected.
Photon energies are calibrated subsequently and their energies in simulated samples are smeared to match the resolution in data \cite{Khachatryan:2014ira}.

Events are required to have at least two identified photon candidates that are within the ECAL and tracker fiducial region ($\abs{\eta} < 2.5$), but excluding the ECAL barrel-endcap transition region ($1.44 < \abs{\eta} < 1.57$). The photon candidates are required to pass the following criteria: \mbox{$100 < \Mgg < 180 \GeV$}; \mbox{$\ptgone/\Mgg > 1/3$} and \mbox{$\ptgtwo/\Mgg > 1/4$}.
In cases where more than two photons are found, the photon pair with the highest transverse momentum $\ptgg$ is chosen.

For events that pass the above selections, the trigger efficiency is measured to be close to 100\% using data events containing a Z boson decaying to a pair of electrons, or to a pair of electrons or muons in association with a photon \cite{Khachatryan:2015iwa}.

\subsection{The \texorpdfstring{$\PH\to \bbbar$}{Higgs to two b quarks} candidate}

The particle-flow (PF) algorithm aims to reconstruct each individual particle (referred to as a PF candidate) in an event with an optimized combination of information from the various elements of the CMS detector~\cite{Sirunyan:2017ulk}.
Jets are clustered from these candidates using the anti-\kt algorithm with a distance parameter $\Rj=0.4$ \cite{AK4, Cacciari:2011ma}.
Jet candidates are required to have $\PT > 25 \GeV$ and $\abs{\eta} < 2.4$. In addition, identification criteria
are applied to remove spurious jets associated with calorimeter noise.
Finally, jets must be separated from each of the two selected photon candidates by a distance $\DRgj \equiv \sqrt{\smash[b]{(\Detagj)^2 +(\Dphigj)^2}}  > 0.4$, where $\phi$ is the azimuthal angle in radians.
The selected jets are combined into dijet candidates.
At least one dijet candidate is necessary for an event to be selected.
The combined secondary vertex algorithm, optimized for 13\TeV data, provides a continuous \PQb tagging score defined between 0 and 1. It is used to quantify the probability that a jet is a result of a \PQb quark hadronization \cite{BTV-16-002}.
In cases where more than two jets are found, the dijet constructed from the two jets with the highest \PQb tagging scores is selected.
 An event is accepted if~\mbox{$70 < \Mjj < 190\GeV$}.

 The energies of the two selected jets are corrected using the standard CMS algorithm, which is flavor blind \cite{JINST6}. In addition to this correction, a jet energy regression procedure is used to improve the $\Mjj$ resolution.
A multivariate analysis technique is used to correct the absolute scale of the heavy-quark jets by taking into account their specific fragmentation features, \ie, a larger contribution from charged leptons and neutrinos than in light-quark jets. The approach we use in this letter is similar to the one used in the CMS search for the SM $\Hbb$ decays~\cite{Sirunyan:2017elk}, and in addition, takes advantage of variables related to the missing transverse momentum vector, \ptvecmiss, to estimate the neutrino contribution to the heavy-quark decay.
The \ptvecmiss is calculated as the negative of the vectorial sum of the transverse momenta of all PF candidates.
The jets forming the dijet candidate are ordered by $\pt$, and an optimized energy regression distinguishes the higher $\pt$ jet from the lower $\pt$ one.
The improvement to the $\Mjj$ resolution due to the regression procedure depends on the \PQb jet $\pt$ spectrum characteristic of a given signal hypothesis. For example, for the SM-like search it is of the order of 15\%.

\subsection{The \texorpdfstring{$\HH$}{HH} system}
\label{HHsystem}

A summary of the baseline selection requirements is presented in Table~\ref{table:gencut}. After the diphoton and dijet candidates are selected, they are combined to form an \HH candidate.

\begin{table}[htb]
\topcaption{Summary of the baseline selection criteria.}
\label{table:gencut}
\centering
\begin{tabular}{l l l l}
\hline
\multicolumn{2}{c}{Photons} & \multicolumn{2}{c}{Jets} \\
Variable & Selection & Variable & Selection \\
\hline \\[-2ex]
$\ptgone $ & ${>}\Mgg/3$ & $\pt$ [\GeVns{}]& ${>}25$ \\
$\ptgtwo $ & ${>}\Mgg/4$ & $\DRgj$ & ${>} 0.4$  \\
$\abs{\eta}$ & ${<}2.5$        & $\abs{\eta}$ & ${<}2.4$              \\
$ \Mgg$ [\GeVns{}] & $[100, 180]$   & $\Mjj$ [\GeVns{}] & $[70,190]$      \\
\hline
\end{tabular}
\end{table}

\begin{figure*}[thb]
  \centering
  \includegraphics[width=0.8\textwidth]{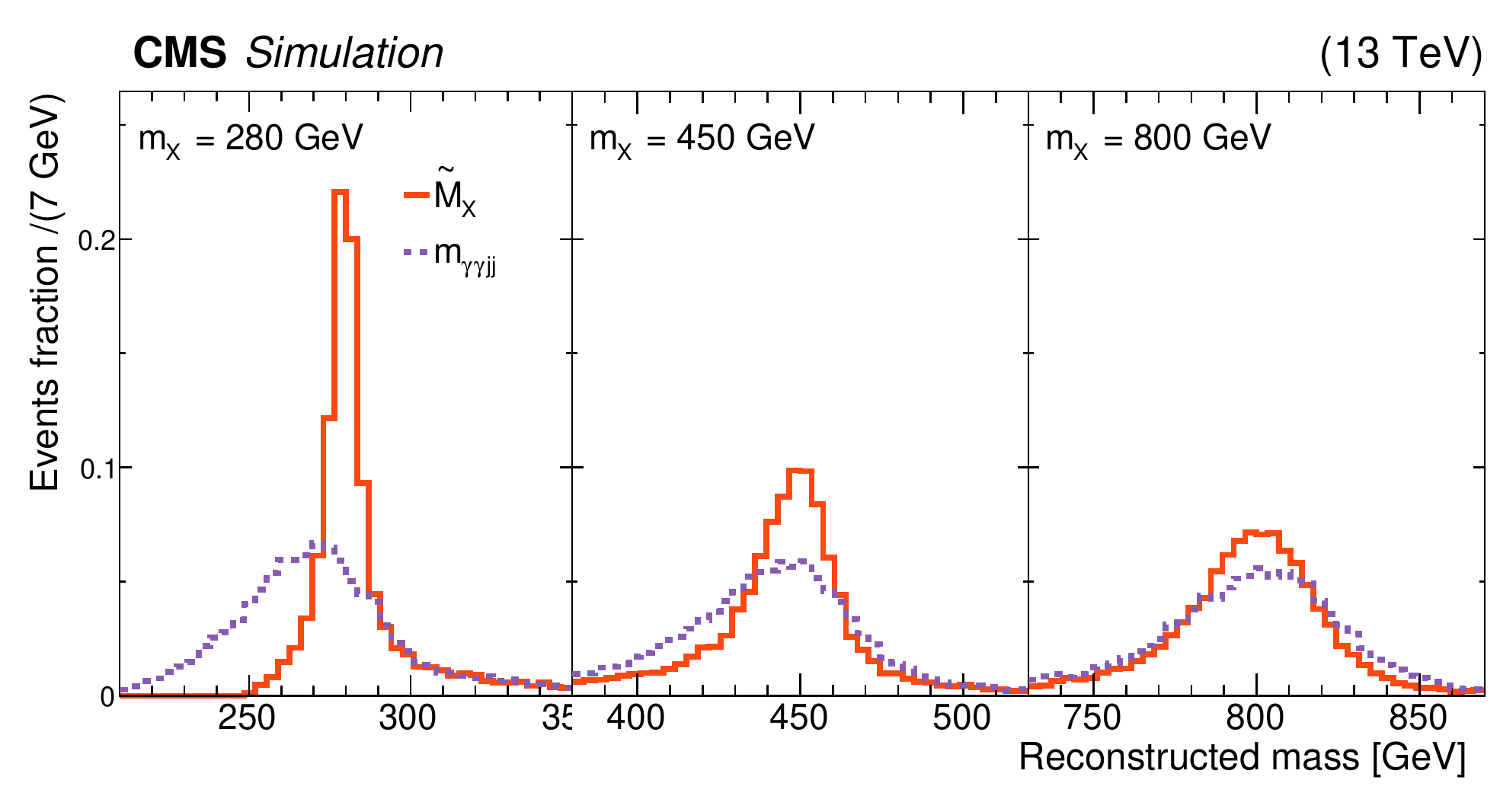}\hfil
  \caption{Comparison of  $\Mtilde$ (red line) with $\Mggjj$ (purple dotted line) for different spin-2 resonance masses. All distributions are obtained after the full baseline selection (Table \ref{table:gencut}), and are normalized to unit area.}
  \label{fig:mx}
\end{figure*}

To have a better estimate of $\mHH$ we correct $\Mggjj$ using
\begin{equation}
\Mtilde = \Mggjj - (\Mjj - \mH) - (\Mgg - \mH),
\end{equation}
which mitigates the $\Mggjj$ dependency on the dijet and diphoton energy resolutions with the assumption that the dijet and diphoton originate from a Higgs boson decay~\cite{Kumar:2014bca}. This procedure has an effect similar to the kinematic fit used previously in Ref.~\cite{Khachatryan:2016sey}. While for the distance parameter $\Rj = 0.5$ used in Ref.~\cite{Khachatryan:2016sey} the kinematic fit was the best option to reconstruct $\mHH$, for the smaller radius used in this paper $\Mtilde$ appears to be more efficient.
The improvements in the $\mHH$ reconstruction are most striking for low-$\mx$ hypotheses, as shown in Fig.~\ref{fig:mx}, and have little impact on high $\mx$ hypotheses. This effect can be understood by the fact that the relative contribution to $\Mtilde$ of $\abs{\Mjj - \mH}$ is much smaller at high $\mx$.

With the four reconstructed objects from the \HH decay, angular correlations in the signal can provide important information to separate it from the background. In this analysis we consider three helicity angles. The scattering angle, $\thetastar$, is defined in the Collins--Soper (CS) frame of the four-body system~\cite{Bolognesi:2012mm},
as the angle between the momentum of the Higgs boson decaying into two photons
and the line that bisects the acute angle between the colliding protons.
Since the directions of the two Higgs boson candidates are collinear in the CS frame, the choice of the Higgs boson decaying to photons as the reference direction is arbitrary. Therefore, we use the
absolute value of the cosine of this angle $\acosthetastar$ to obviate this arbitrariness.
The \PH boson decay angles are defined, in a way similar to Ref. \cite{Bolognesi:2012mm}, as the angles of the decay products in each Higgs boson's rest frame with respect to the direction of motion of the boson in the CS frame. Since the two photons from the Higgs boson decay are indistinguishable and the charges of the \PQb quarks are not considered in this analysis, the absolute values of the cosines of these angles are used:  $\acosthetagg$ and $\acosthetabb$.

\begin{figure*}[hbt!]
\centering
\includegraphics[width=0.35\textwidth]{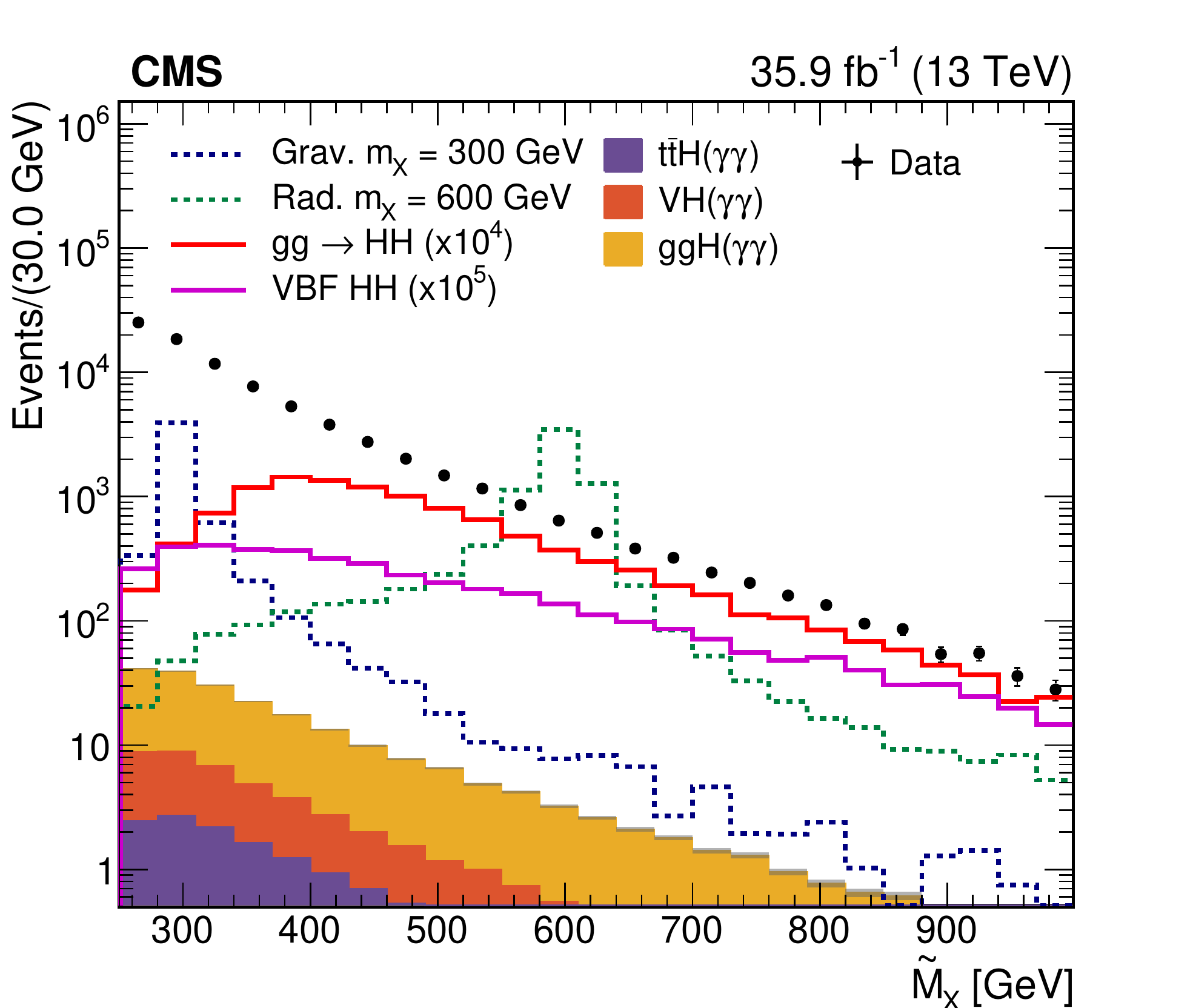}
\includegraphics[width=0.35\textwidth]{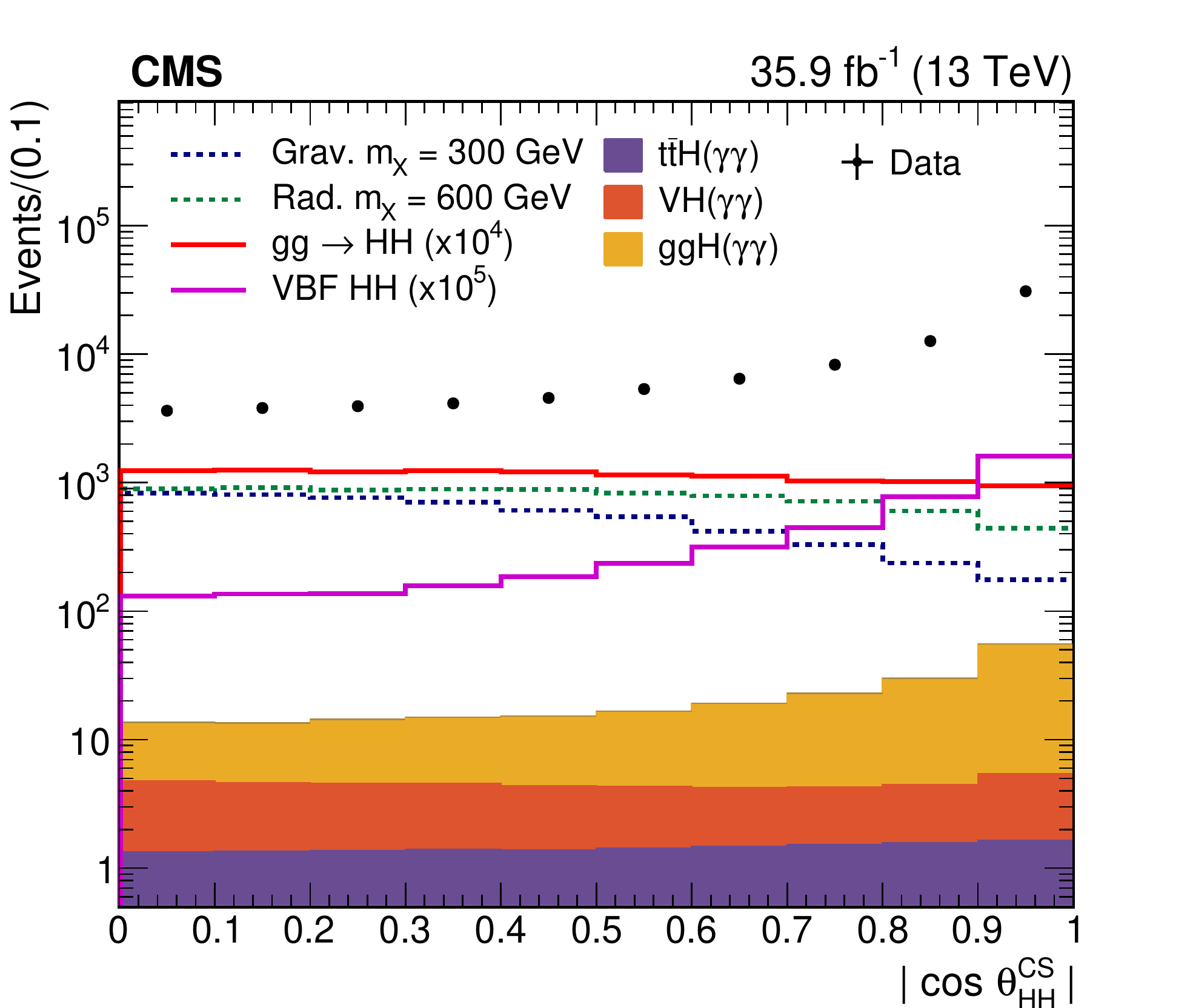}
\\[0.5cm]
\includegraphics[width=0.35\textwidth]{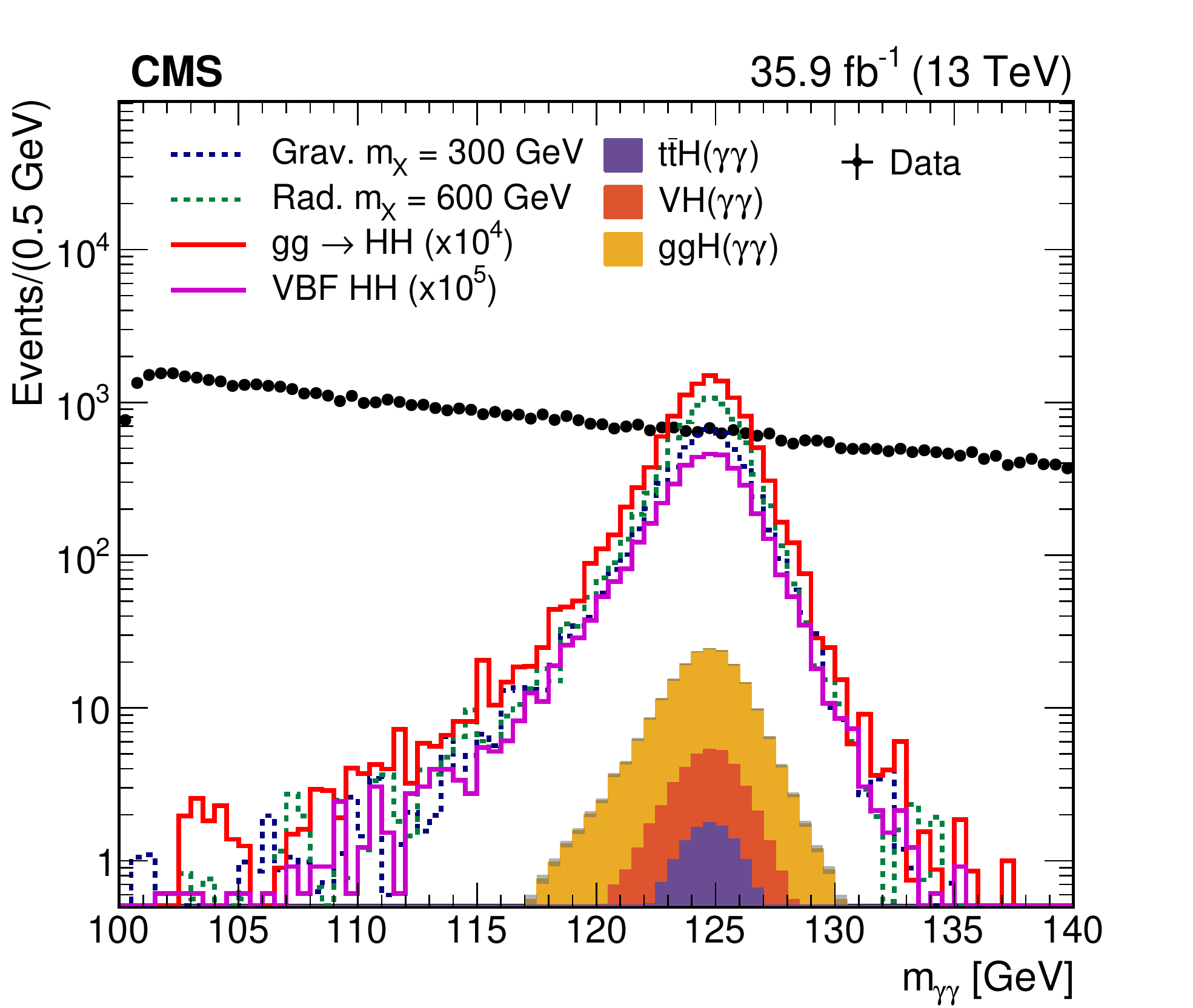}
\includegraphics[width=0.35\textwidth]{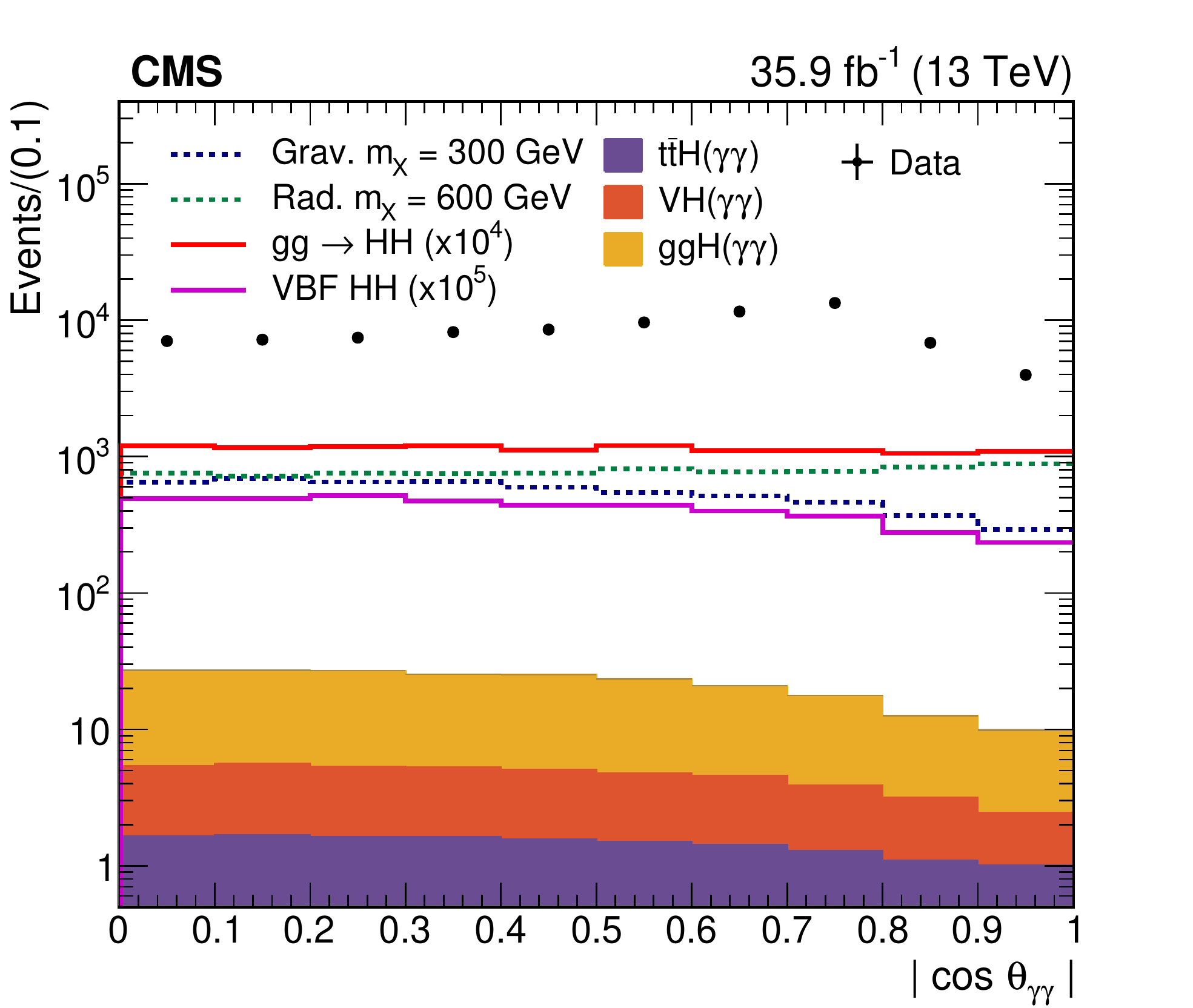}
\\[0.5cm]
\includegraphics[width=0.35\textwidth]{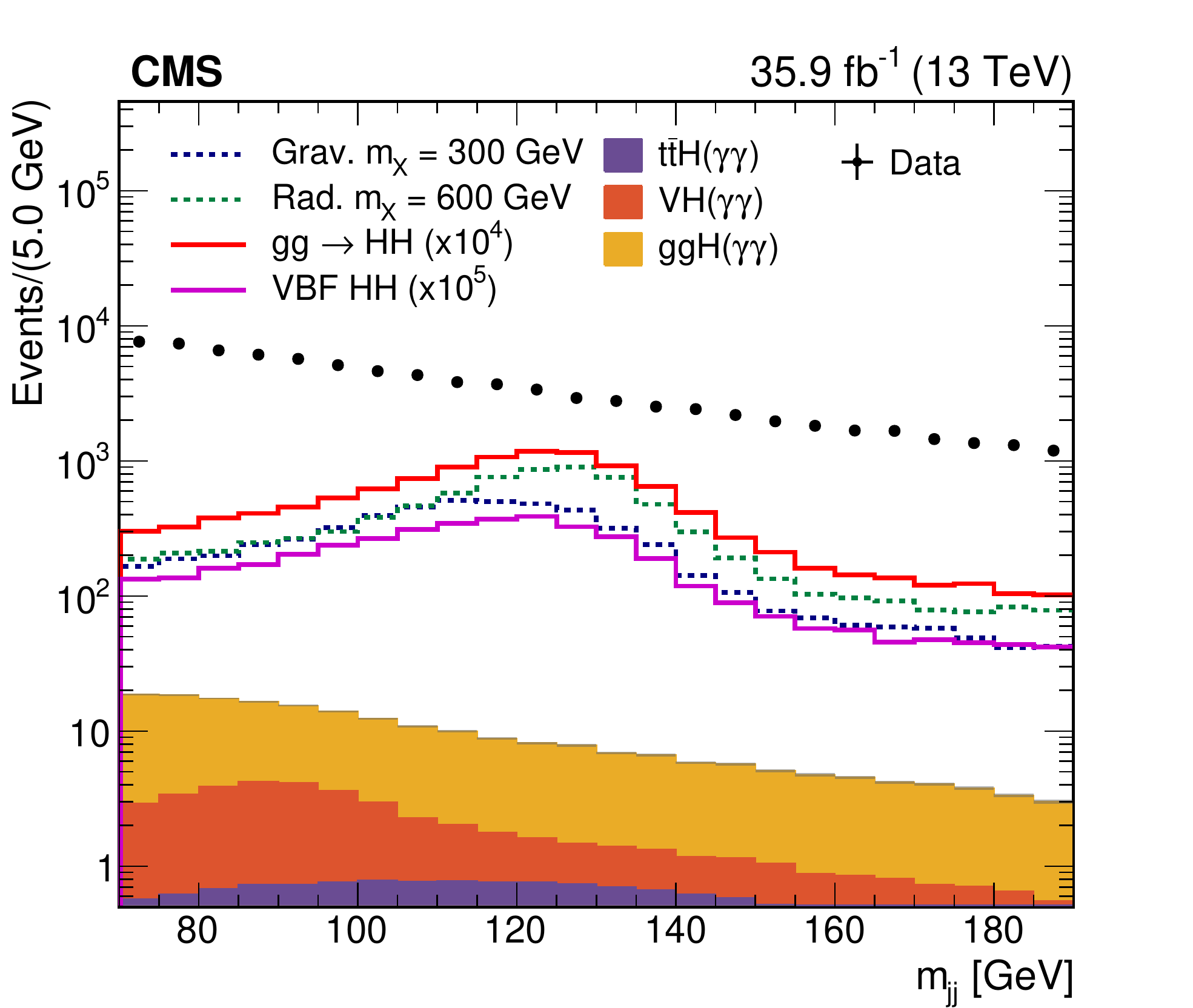}
\includegraphics[width=0.35\textwidth]{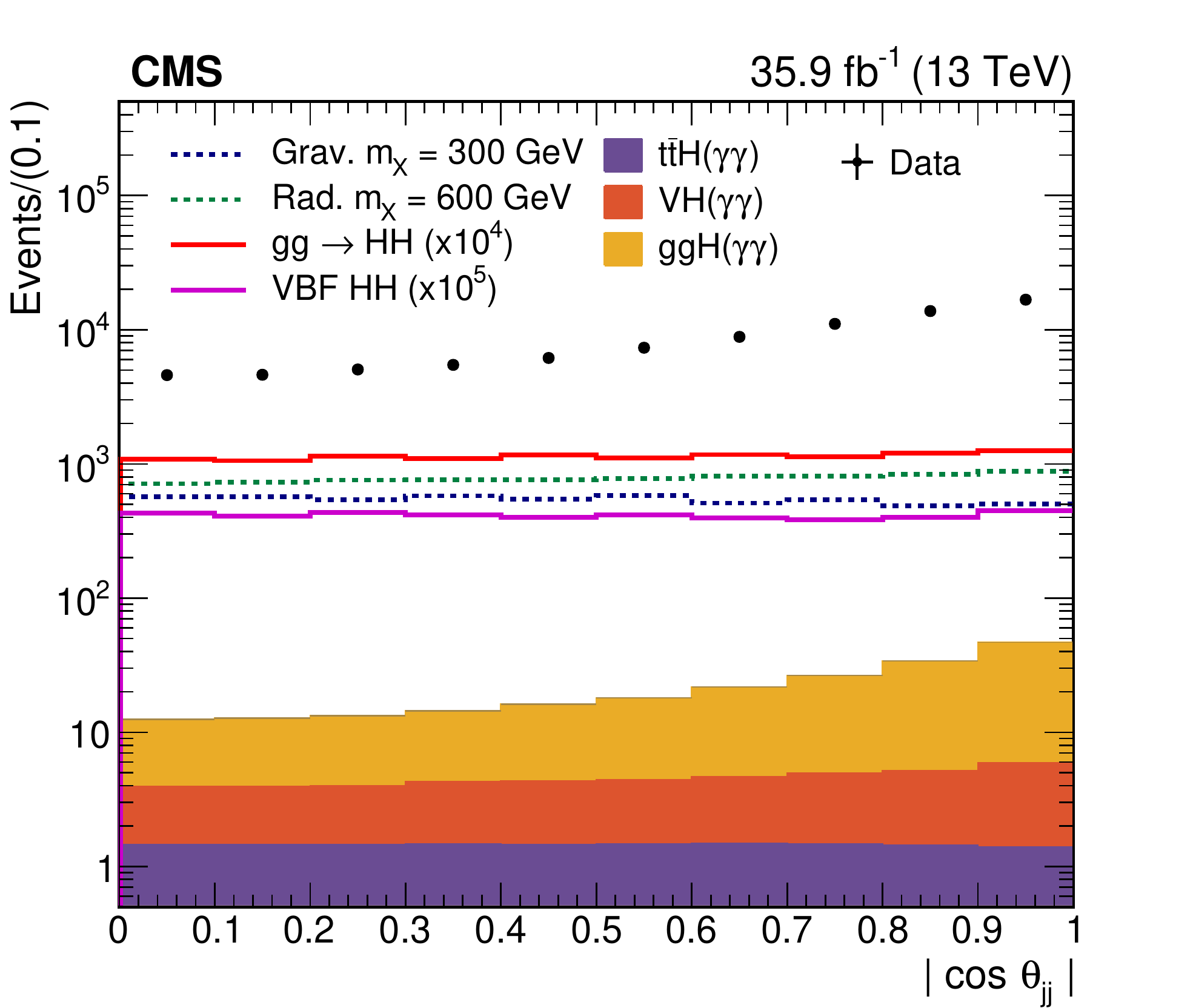}

\caption{Data (dots), dominated by \QCD background, compared to different signal hypotheses and three single-Higgs boson samples (\ttH, \VH, and \ggH) after the selections on photons and jets summarized in Table \ref{table:gencut} for the kinematic distributions described in Sections \ref{sec:intro} and \ref{HHsystem}: $\Mtilde$ (top left) and $\acosthetastar$ (top right); $\Mgg$ (middle left) and $\acosthetagg$ (middle right); $\Mjj$ (bottom left) and $\acosthetabb$ (bottom right). The statistical uncertainties on the data are barely visible beyond the markers. The resonant signal cross section is normalized to 500\unit{fb} and the SM-like \ggHH (VBF \HH) process to $10^4$ ($10^5$) times its cross section.}
\label{figure:control_plots}\end{figure*}

\subsection{Background properties}
\label{section:background}

The main kinematic distributions of the $\bbgg$ final state that are used throughout the analysis (invariant masses and helicity angles) are shown in Fig. \ref{figure:control_plots}, after the basic selections summarized in Table \ref{table:gencut}.
The data in Fig. \ref{figure:control_plots} are dominated by \QCD events, which are the primary contribution to the background in this region of phase space.
The SM single-Higgs boson production processes, represented by colored areas in the figure, are three orders of magnitude lower than the nonresonant \QCD processes. Only single-Higgs boson production processes with a
sufficient number of events (\ttH, \VH, and \ggH) are shown for clarity of the figure.
Finally, signal shapes are shown in the figures, where the resonant ones have been normalized to a cross section of 500\unit{fb} and the SM-like \ggHH (VBF \HH) signal to $10^4$ ($10^5$) times its cross section.

As expected, the signals produce peaks in $\Mgg$ and $\Mjj$. The resonant di-Higgs boson signals peak sharply in $\Mtilde$, while SM \HH processes exhibit broad structures induced by the interference pattern of different Feynman diagrams contributing to the $\HH$ production and shaped by the analysis selections. The data show a smoothly falling mass spectrum, as expected for the \QCD background. Finally, the single-Higgs boson backgrounds peak in $\Mgg$, but not in $\Mjj$ or $\Mtilde$ (except the V$\PH$ process which gives a peak in $\Mjj$ around the V masses).

The $\acosthetastar$ distribution is sensitive to the tensor structure of the production mechanism (see for example Ref.~\cite{Chizhov:2011wt}).
It is relatively flat for $\ggHH$ \cite{Azatov:2015oxa} and the spin-0 mediated production. For the spin-2 mediated production it decreases toward 1, while for VBF \HH and the data it rises toward 1.
The distribution of the cosine of the $\Hgg$ helicity angle is expected to be flat for the samples with genuine Higgs bosons. The decrease toward 1 is due to the selections on photon $\pt$.
In the data the distribution rises up to 0.8 and then decreases. This shape results from the combination of matrix element properties and the asymmetric selections on the photon $\pt$.
In the same way, the $\acosthetabb$ distribution is flat for the signal, but rises significantly toward 1 for the data and \ggH.

\section{Event classification and modeling}
\label{sec:AnalysisMethods}

After dijet and diphoton candidate selection, events are placed into categories using the $\Mtilde$ variable and a multivariate (MVA) classifier. Both variables are designed to minimize the correlation between $\Mgg$ and $\Mjj$.
In each category, a parametric fit is performed in the two-dimensional $\Mgg$\,--\,$\Mjj$ plane for the signal extraction procedure using a product of probability densities (PDs) for signal and backgrounds. This 2D approach helps to constrain the impact of the single-Higgs boson background since its structure in $\Mjj$ differs from that of the signal. Finally, all the categories are combined together assuming a signal model to maximize the sensitivity of the analysis.

\subsection{Event classification}

\subsubsection{\texorpdfstring{$\Mtilde$}{M-tilde} categorization}

In the nonresonant search, a categorization is performed using the $\Mtilde$ information.
Since the $\Mtilde$ spectrum for SM-like $\ggHH$ production has a maximum at around 400\GeV and the \QCD background peaks at the kinematic threshold of $250\GeV = 2\mH$, the maximal sensitivity is achieved for $\Mtilde > 350\GeV$.  However, anomalous couplings may change the $\Mtilde$ distribution for the signal hypothesis. Therefore, instead of imposing a $\Mtilde$ selection, events are categorized in the nonresonant search into high-mass (HM) and low-mass (LM) regions that are above and below $\Mtilde = 350\GeV$, respectively.

In the resonant search, $\Mtilde$ is used to define a unique signal region that depends on the mass of the resonance being sought. This mass window typically contains 60\% of the signal at low $\mx$, increasing gradually for higher $\mx$. The resonant search starts just above the threshold at $260\GeV \gtrapprox 2\mH$ and extends up to $\mx = 900\GeV$.
In fact the $\Rj$ value used in this paper is small enough to reconstruct the decay products of two boosted \PQb quarks produced in the Higgs boson decays as separate jets up to $\mx \approx 1.25\TeV$~\cite{Gouzevitch:2013qca}. However, for values of $\mx \approx 1\TeV$ and larger the available amount of data is too small to perform the signal extraction procedure as defined in this paper.

\subsubsection{MVA categorization}

An MVA procedure is used to select the most signal-like events and to further classify them.
With this goal, a boosted decision tree (BDT) is trained with the {\sc TMVA} package~\cite{Hocker:2007ht} using three types of variables:

\begin{itemize}
\item \PQb tagging variables: the \PQb tagging score of each jet in the dijet candidate;
\item Helicity angles as defined in Section \ref{HHsystem};
\item \HH transverse balance variables: $\ptgg/\Mggjj$ and $\ptjj/\Mggjj$, where $\ptjj$ is the transverse momentum of the dijet candidate.
\end{itemize}

The BDT is trained with the ensemble of \ggHH samples as the signal hypothesis in the nonresonant search separately for low- and high-mass categories. For the resonant cases, the ensemble of resonant signals is used to train one BDT for $\mx < 600\GeV$ and another one for $\mx > 600\GeV$. This training strategy maximizes the sensitivity to massive resonances.
The background events used for the training are obtained from a control sample that was extracted from the data by inverting the identification condition on one of the two photons. This sample is dominated by events having a photon produced with three accompanying jets.
We verified that after excluding the events with $120 < \Mgg < 130\GeV$ in the signal and control samples, the kinematic properties of these two samples are well matched.

\begin{figure}[thb]
  \centering
  \includegraphics[width=0.49\textwidth]{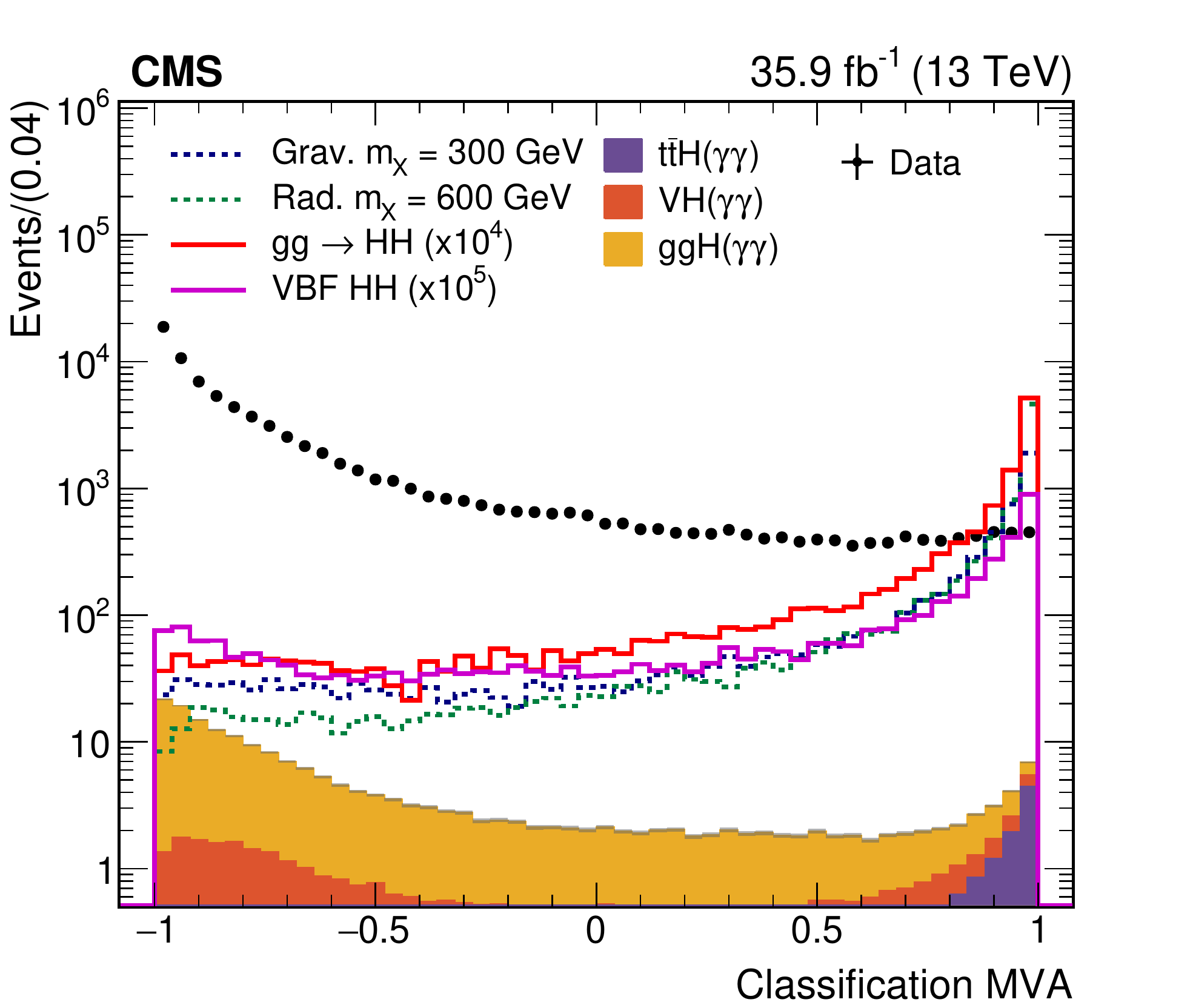}\hfil
  \caption{Distributions of the BDT output (classification MVA) obtained for the high-mass nonresonant training.
Data, dominated by \QCD background, are compared to different signal hypotheses and three single-Higgs boson samples (\ttH, \VH, and \ggH) after the selections on photons and jets summarized in Table \ref{table:gencut}. The statistical uncertainties on the data are barely visible beyond the markers. The resonant signal cross section is normalized to 500\unit{fb} and the SM-like \ggHH (VBF \HH) process to $10^4$ ($10^5$) times its cross section.}
  \label{fig:hhtagger}
\end{figure}

Figure~\ref{fig:hhtagger} shows the BDT output from one of the four trainings. The MVA efficiently separates gluon-gluon produced signals from the \QCD background that represents the dominant contribution to the data.
The most powerful discriminating variables used in the BDT are the b tagging scores of the jets, followed by the kinematic variables. Therefore, the single-Higgs boson production samples with genuine contributions from two \PQb quarks (\ttH and $\PZ(\to\bbbar)\PH$) are classified as more signal-like, while \ggH and other \VH processes are classified as more background-like.
Finally, events from the VBF \HH production are selected less efficiently than those from \ggHH production.

\begin{table*}[htb]
\centering
\topcaption{Definition of high-purity category (HPC) and medium-purity category (MPC) for the resonant and nonresonant analyses.}
\begin{tabular}{ l  l l l}
\hline
Analysis & Region    & Classification MVA & $\Mtilde$           \\ \hline
\multirow{2}{*}{Nonresonant} & High-mass & \begin{tabular}[l]{@{}l@{}}HPC: $\text{MVA} > 0.97$\\ MPC: $0.6 < \text{MVA} < 0.97$\end{tabular} & $\Mtilde > 350\GeV$ \\[\cmsTabSkip]
                             & Low-mass  & \begin{tabular}[l]{@{}l@{}}HPC: MVA \textgreater~ 0.985\\ MPC: 0.6 \textless~ MVA \textless~ 0.985\end{tabular} & $\Mtilde < 350\GeV$ \\[\cmsTabSkip]

\multirow{2}{*}{Resonant}    & $\mx > 600\GeV$ & \begin{tabular}[l]{@{}l@{}}  HPC: MVA \textgreater~ 0.5\\ MPC: 0 \textless~ MVA \textless~ 0.5\end{tabular} & Mass window         \\[\cmsTabSkip]
& $\mx < 600\GeV$  & \begin{tabular}[l]{@{}l@{}}HPC: MVA \textgreater~ 0.96\\ MPC: 0.7 \textless~ MVA \textless~ 0.96\end{tabular} & Mass window\\
\hline
\end{tabular}
\label{tab:cats}
\end{table*}

For a given category the purity is defined as the ratio between the number of events coming from a hypothetical signal with a production cross section normalized to 1\unit{fb} and the number of background events.
For each of the four trainings, the output of the MVA classifier is used to define a category with the highest purity (HPC) and another with medium purity (MPC). The remaining events are rejected, because they do not improve the sensitivity of the analysis.
In the nonresonant low-$\Mtilde$ MPC region, an additional requirement is placed on the \PQb tagging score, corresponding to 80\% efficiency for genuine $\PQb$ jets \cite{BTV-16-002}.
This reduces the contribution of the events where the jet with lowest $\PQb$ tagging score comes from a pileup event.
Table \ref{tab:cats} shows the HPC and MPC definitions for the different regions of the resonant and nonresonant analyses.

\subsubsection{Signal acceptance times efficiency}

\begin{figure}[thb!]
  \centering
  \includegraphics[width=0.49\textwidth]{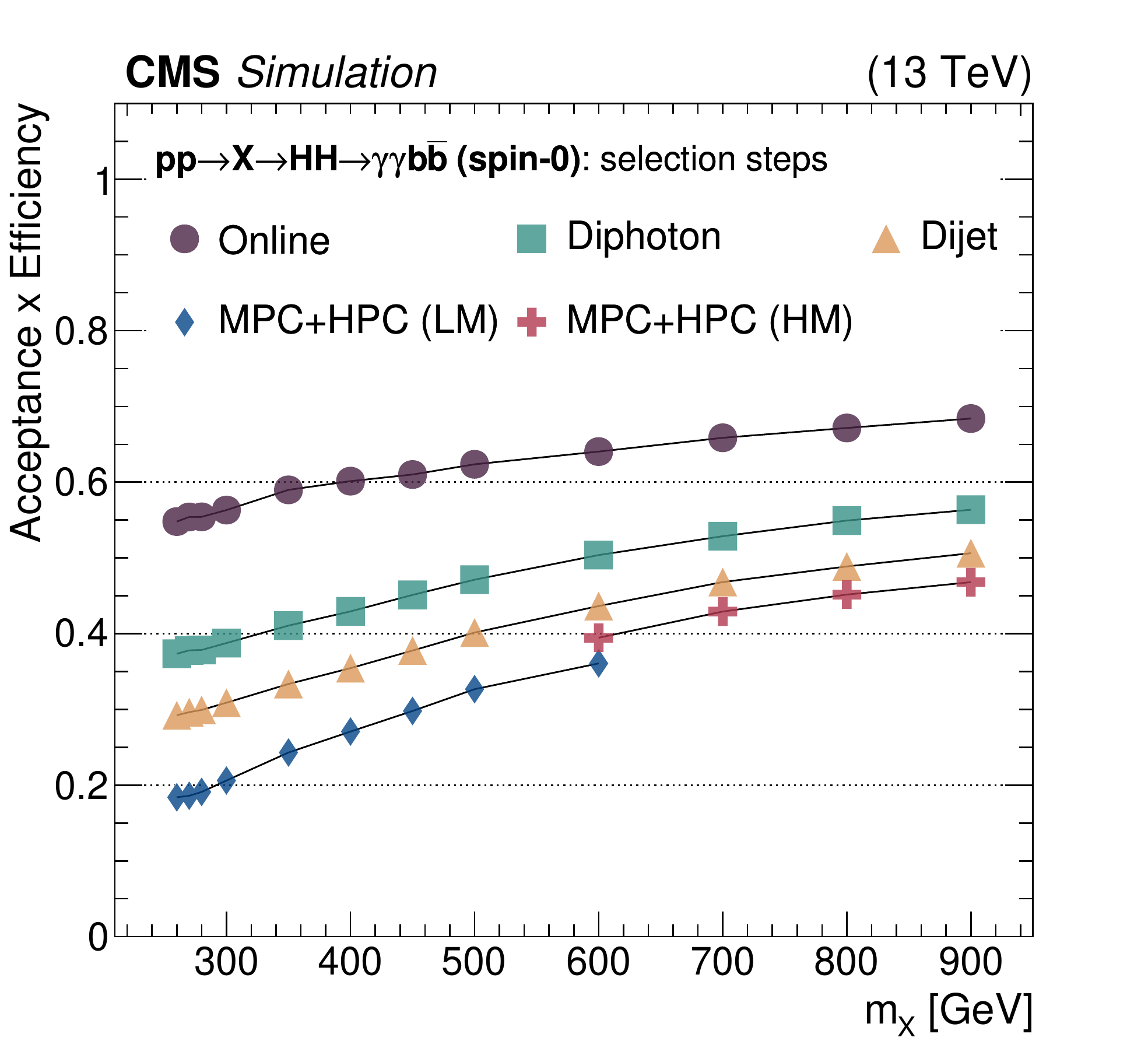}\hfil
  \includegraphics[width=0.49\textwidth]{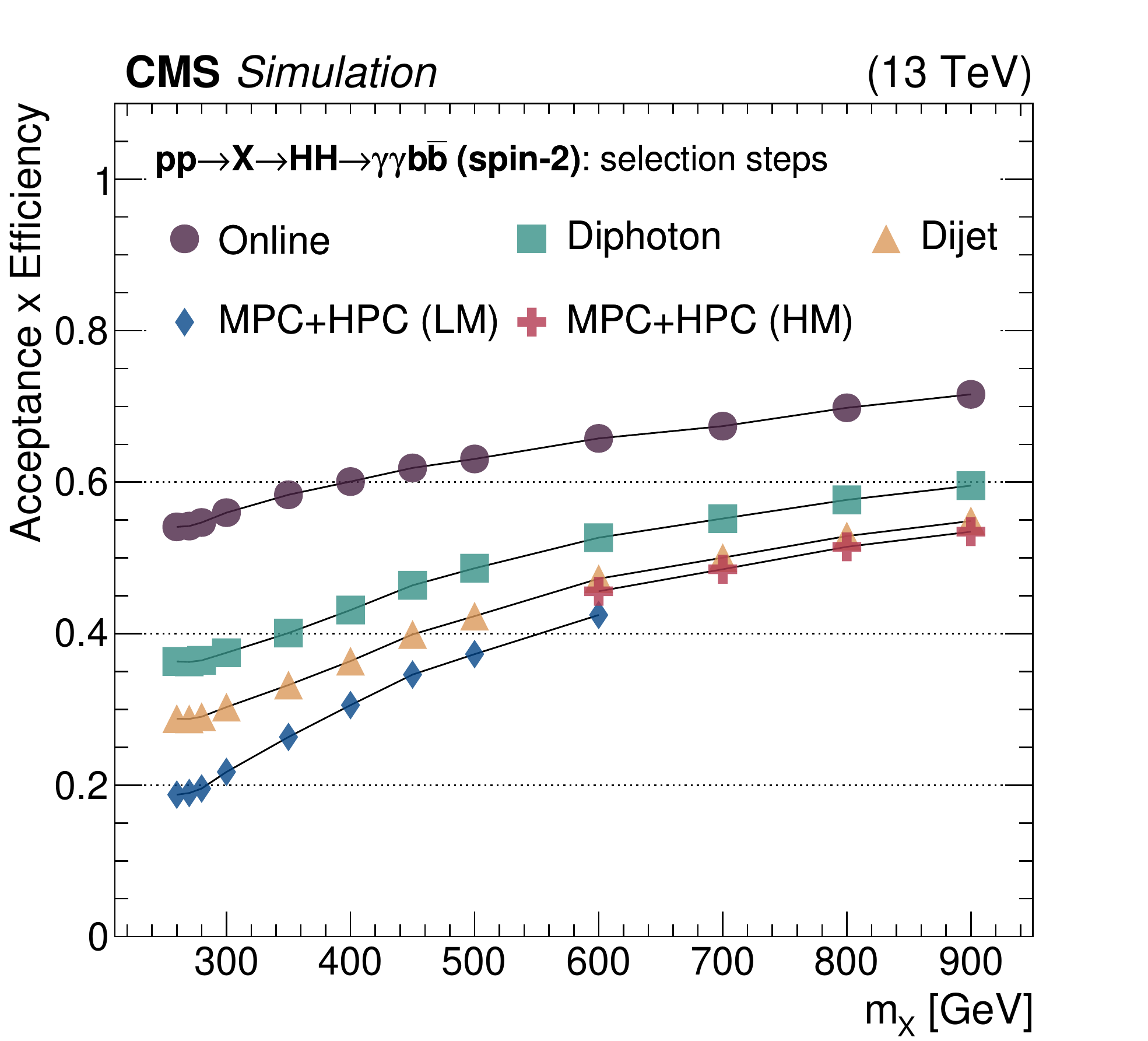}\hfil
  \caption{Consecutive selection efficiencies for different analysis steps for two resonance hypotheses: spin-0 (\cmsLeft) and spin-2 (\cmsRight). Online selection includes the photon online preselection conditions described at the beginning of Section \ref{sec:objects}. Diphoton selections include photon identification and kinematics selections from Table \ref{table:gencut}. Dijet selections are those described in Table \ref{table:gencut}.}
  \label{fig:cutflow-signal}
\end{figure}

The typical signal acceptance times efficiency (\AxE) values obtained in the resonant analysis are shown in Fig.~\ref{fig:cutflow-signal}. Consecutive effects of the selections on \AxE for different analysis steps are shown: after the trigger selection that includes a loose online preselection on the photons, after the diphoton candidate selection, after the dijet candidate selection, and after the MVA categorization.
The final \AxE values range from approximately 20\% (low mass) to 50\% (high mass) for both the spin-0 and spin-2 resonance hypotheses. For an identical $\mx$ value, \AxE is a bit higher for a spin-2 hypothesis than for spin-0, because the Higgs bosons are, on average, produced more centrally in the KK graviton model considered in this letter (see Fig. \ref{figure:control_plots}).

The \AxE value is 30 (13)\% for the SM-like \ggHH (VBF \HH) signal hypothesis,
with 25 (10)\% in the high-mass region and 5 (3)\% in the low-mass one. The difference between the two production mechanisms mainly comes from the fact that the MVA was trained assuming a \ggHH signal. For example, one of the most discriminating variables, $\acosthetastar$, shown in Fig.~\ref{figure:control_plots}, has a similar behavior to the \QCD background and for the VBF \HH process, while it is very different for the \ggHH process.

\subsection{Signal modeling}
\label{sec:modeling}

The signal PD of each mass dimension
is modeled with a double-sided Crystal Ball (CB) function, which is a modified version of the standard CB function~\cite{CrystalBallRef} with two independent exponential tails.
This modeling is useful in situations in which a lower-energy tail might be created by energy mismeasurements and a higher-energy tail by the mismatching of objects (for example when a jet from additional QCD radiation is misidentified as one of the jets from the \PH boson decay).
The final two-dimensional signal model PD is the product of the independent $\Mgg$ and $\Mjj$ models.
The no-correlation hypothesis is checked by comparing the two-dimensional $\Mgg$\,--\,$\Mjj$ distribution from the simulated signal samples with the two-dimensional PD built as a product of one-dimensional ones.
For the typical expected number of signal events in this analysis, the impact of such correlations is found to be negligible.

\begin{figure}[thb!]
  \centering
  \includegraphics[width=0.49\textwidth]{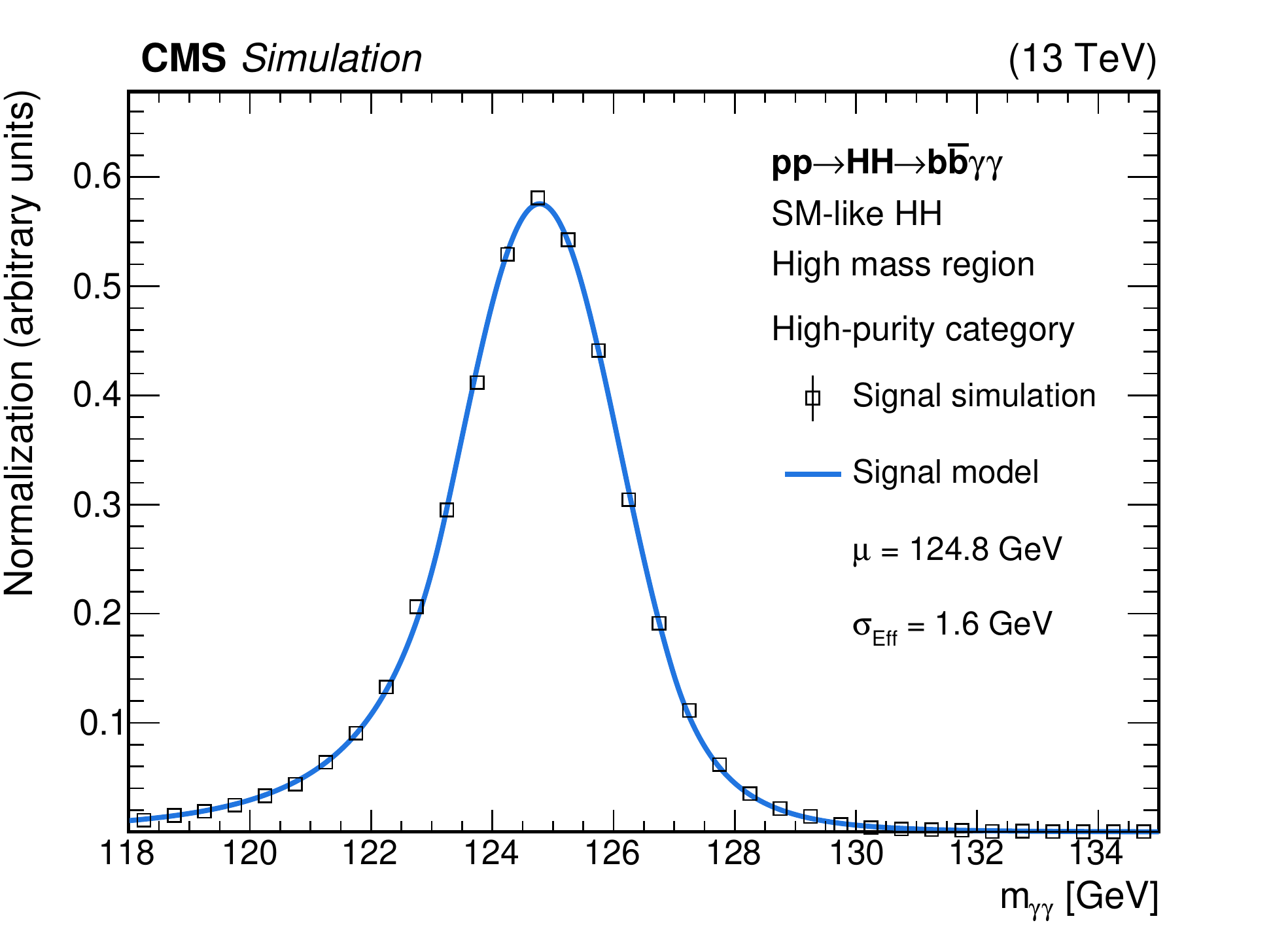}\hfil
  \includegraphics[width=0.49\textwidth]{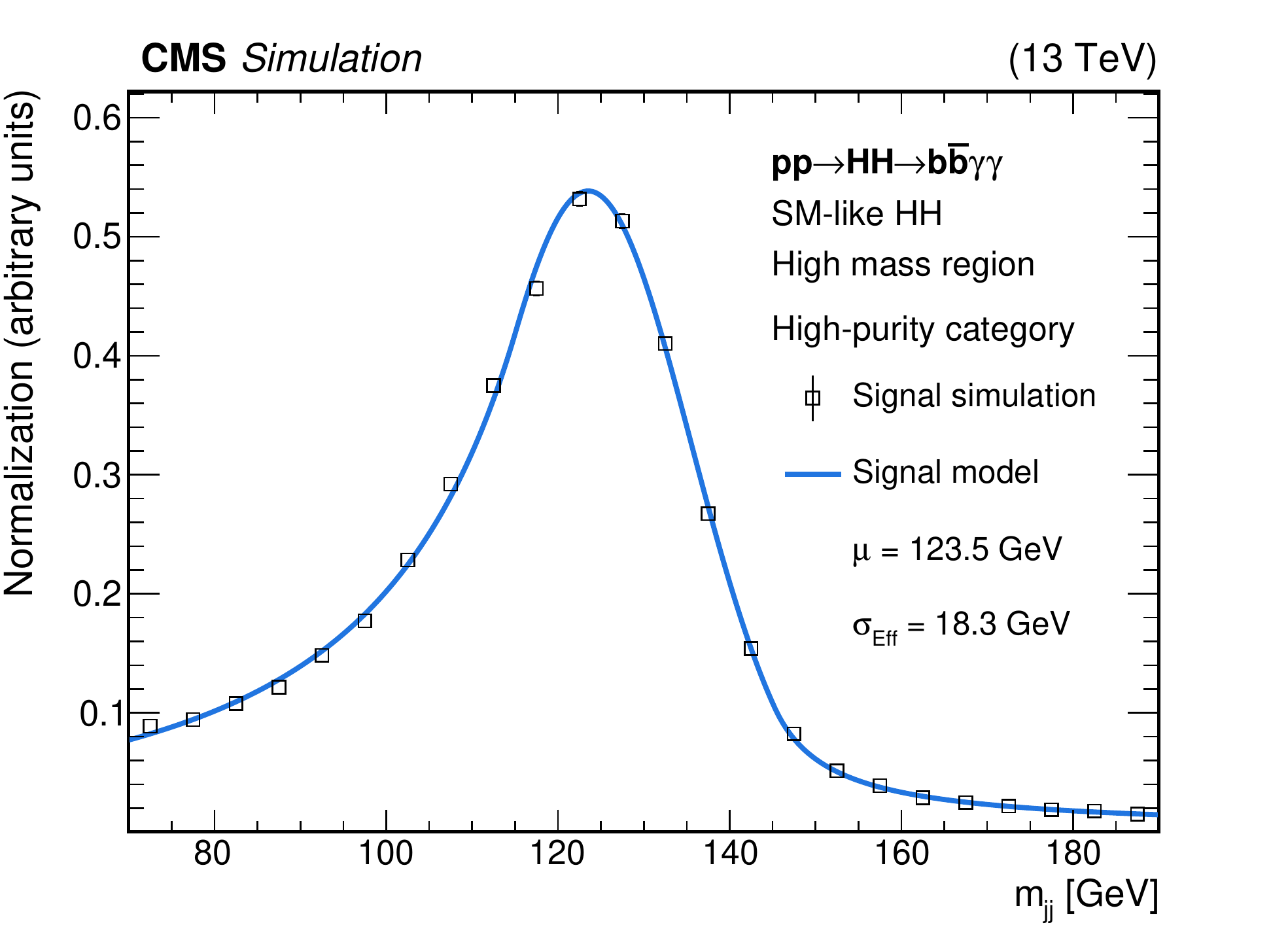}\hfil
  \caption{
Signal shapes for $\Mgg$ (\cmsLeft) and $\Mjj$ (\cmsRight) in the SM \HH nonresonant sample after full analysis selection in the high-mass and HPC region. The solid line shows a fit to the simulated data points with a double-sided Crystal Ball function. The normalization of the shapes is arbitrary.}
  \label{fig:sig_highmassSM}
\end{figure}

The PD parameters are obtained by fitting the simulated signal samples in each analysis region. For each $\mx$ point and each nonresonant sample a dedicated fit is performed.
The resolution is estimated by the $\sigma_\text{eff}$ value, defined as half of the width of the narrowest region containing 68.3\% of the signal shape.
Examples of the signal shapes in the nonresonant analysis, assuming an SM-like signal, are shown in Fig.~\ref{fig:sig_highmassSM}. The diphoton resolution is determined to be ${\approx} 1.6\GeV$ and the dijet resolution is ${\approx} 20\GeV$. The mean of the Gaussian core of the CB function, $\mu$, is close to $\mH$, within 0.1-0.2\% for $\Mgg$ and 1-2\% for $\Mjj$.

\subsection{Background modeling}

The total background model is obtained as a sum of the \QCD background continuum PD and single-Higgs boson production PDs, the latter being normalized to their SM production cross sections.

For both the resonant and nonresonant analyses the \QCD continuum is described using polynomials in the Bernstein basis \cite{Khachatryan:2014ira}.
The data control sample described in Section \ref{sec:modeling} is used to define the appropriate order of the polynomial function for the one-dimensionnal PDs of the \QCD background continuum. For each search we randomly select from the control sample a number of events equal to the total number of events observed in data.
A second-order Bernstein polynomial fits the data well. In categories with fewer events, occurring in the resonant analysis, a first-order Bernstein polynomial is used.
This choice of the background PD is tested for possible biases in the signal extraction by comparing it to other possible background models, such as exponentials and Laurent polynomials.
The bias from the chosen PD is always found to be  smaller than the statistical uncertainty in the
fit, and can be safely neglected~\cite{Chatrchyan:2012ufa}.
The correlation between $\Mgg$ and $\Mjj$ was measured in the data control sample and found to be compatible with zero.

\begin{figure*}[htb]
  \centering
  \includegraphics[width=\plotsize\textwidth]{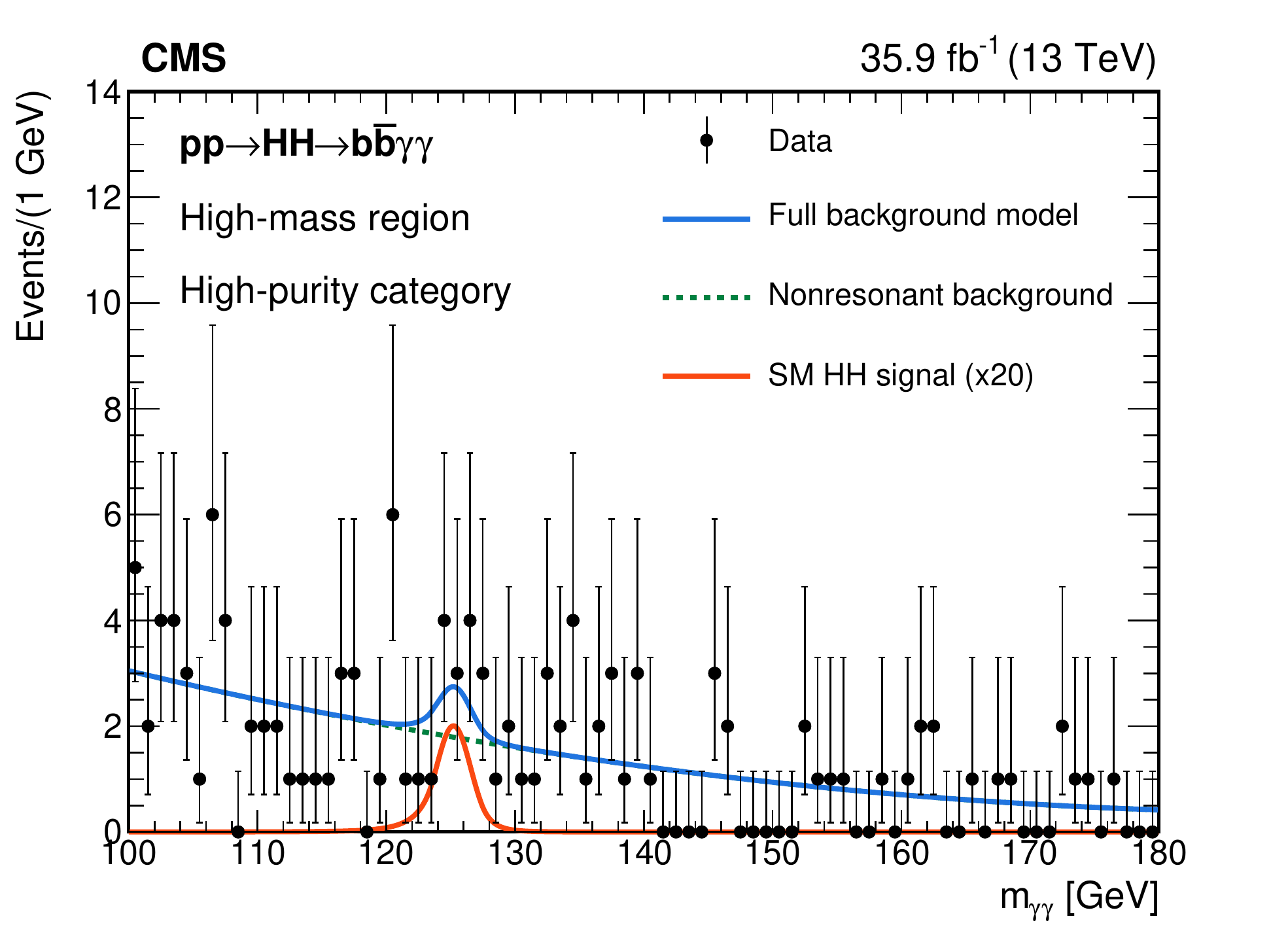}\hfil
  \includegraphics[width=\plotsize\textwidth]{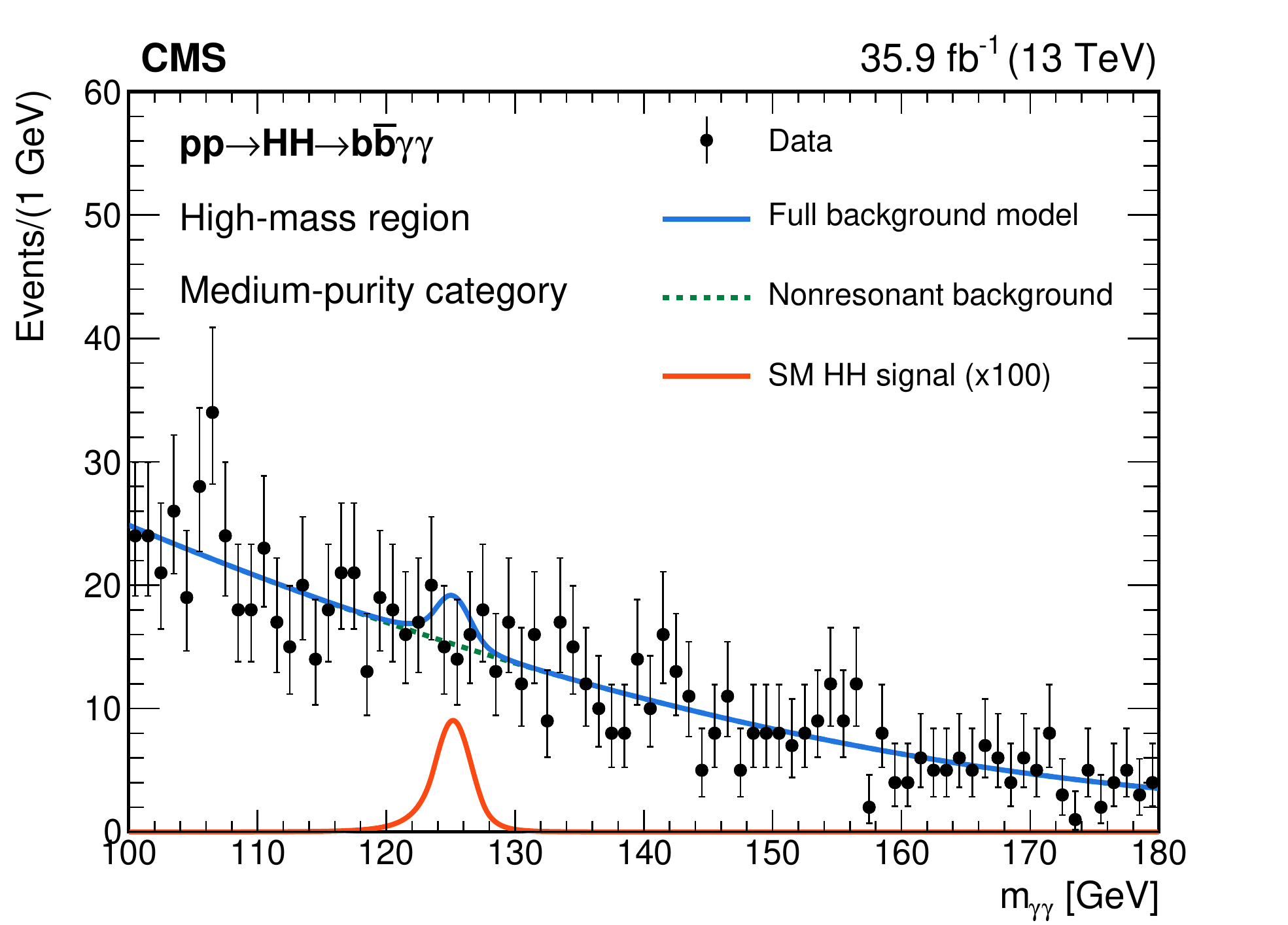}\hfil
  \includegraphics[width=\plotsize\textwidth]{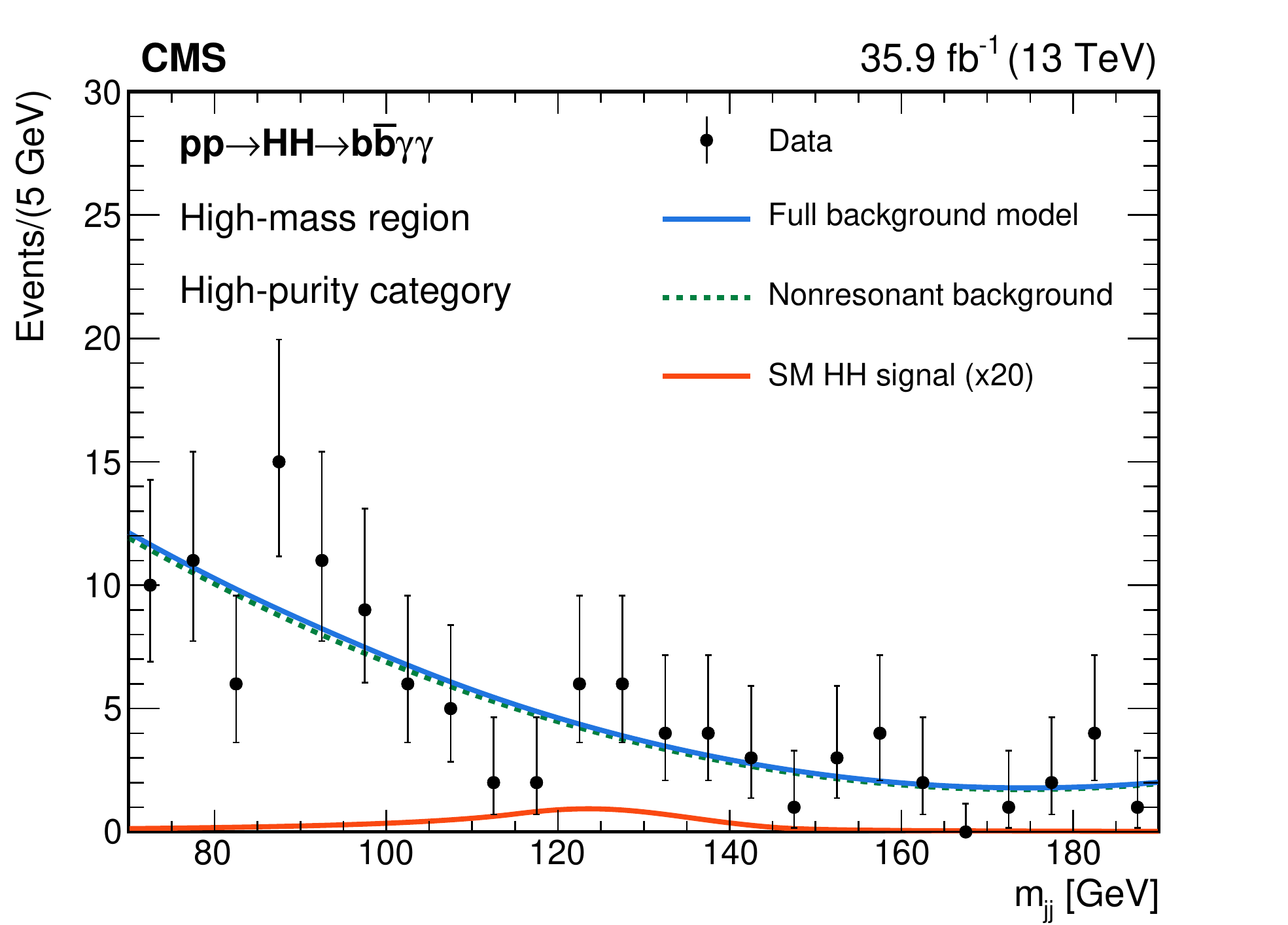}\hfil
  \includegraphics[width=\plotsize\textwidth]{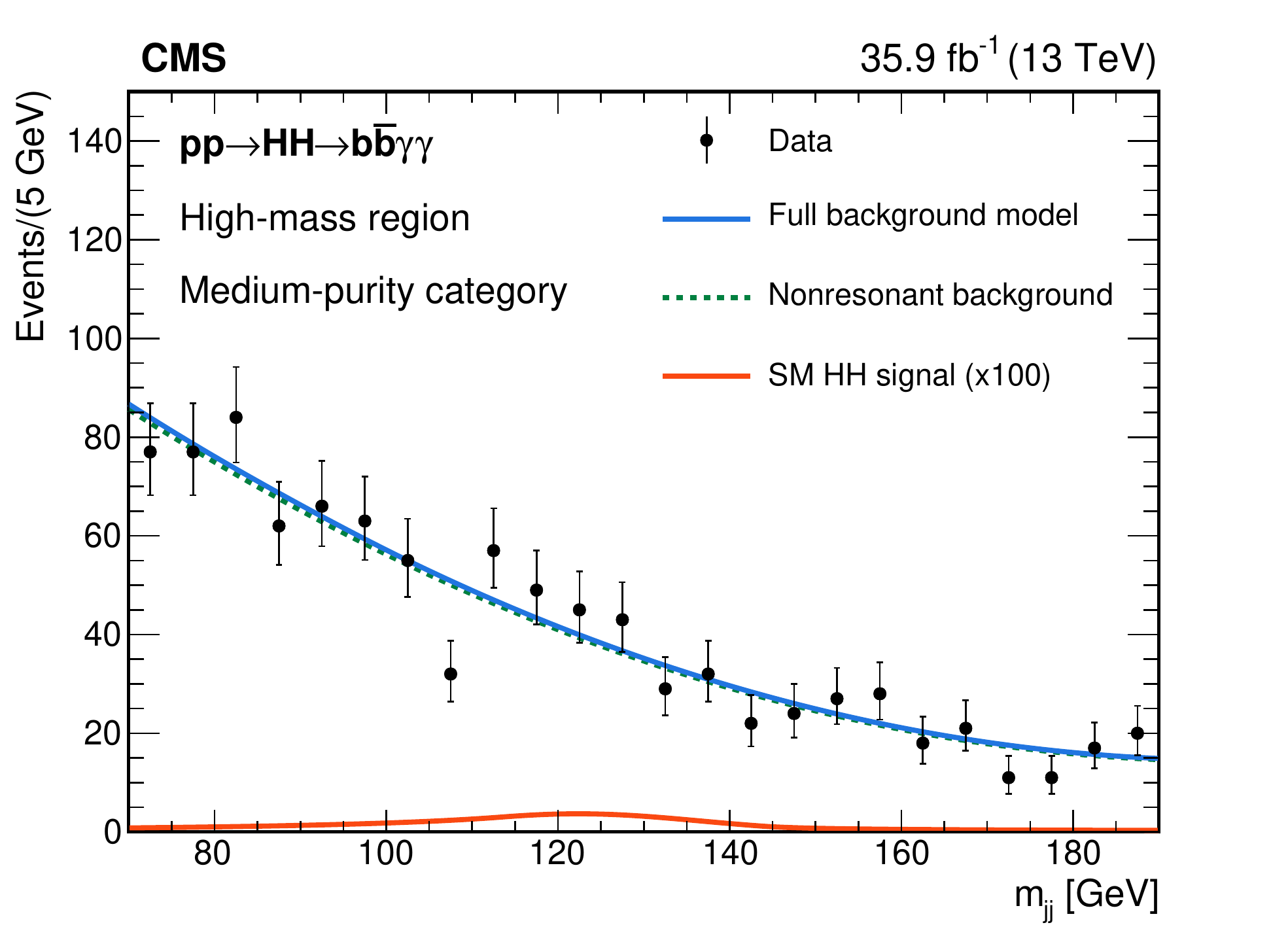}\hfil
  \caption{Background fits for the SM \HH nonresonant analysis selection in the HM region. The plots on the left (right) show the distributions in the HPC (MPC) region. Top plots show the $\Mgg$ spectra and bottom ones $\Mjj$ ones.
The green dashed line represents the nonresonant part of the expected background; the solid blue line represents the full background modeling PD, \ie, the sum of nonresonant background and SM single-Higgs boson contributions scaled to their cross sections;
and the solid red line represents the SM-like \HH production, normalized to its SM cross section times a scaling factor specified in the legend.}
  \label{fig:bkg_fit_nonres_0}
\end{figure*}

\begin{figure*}[htb]
  \centering
  \includegraphics[width=\plotsize\textwidth]{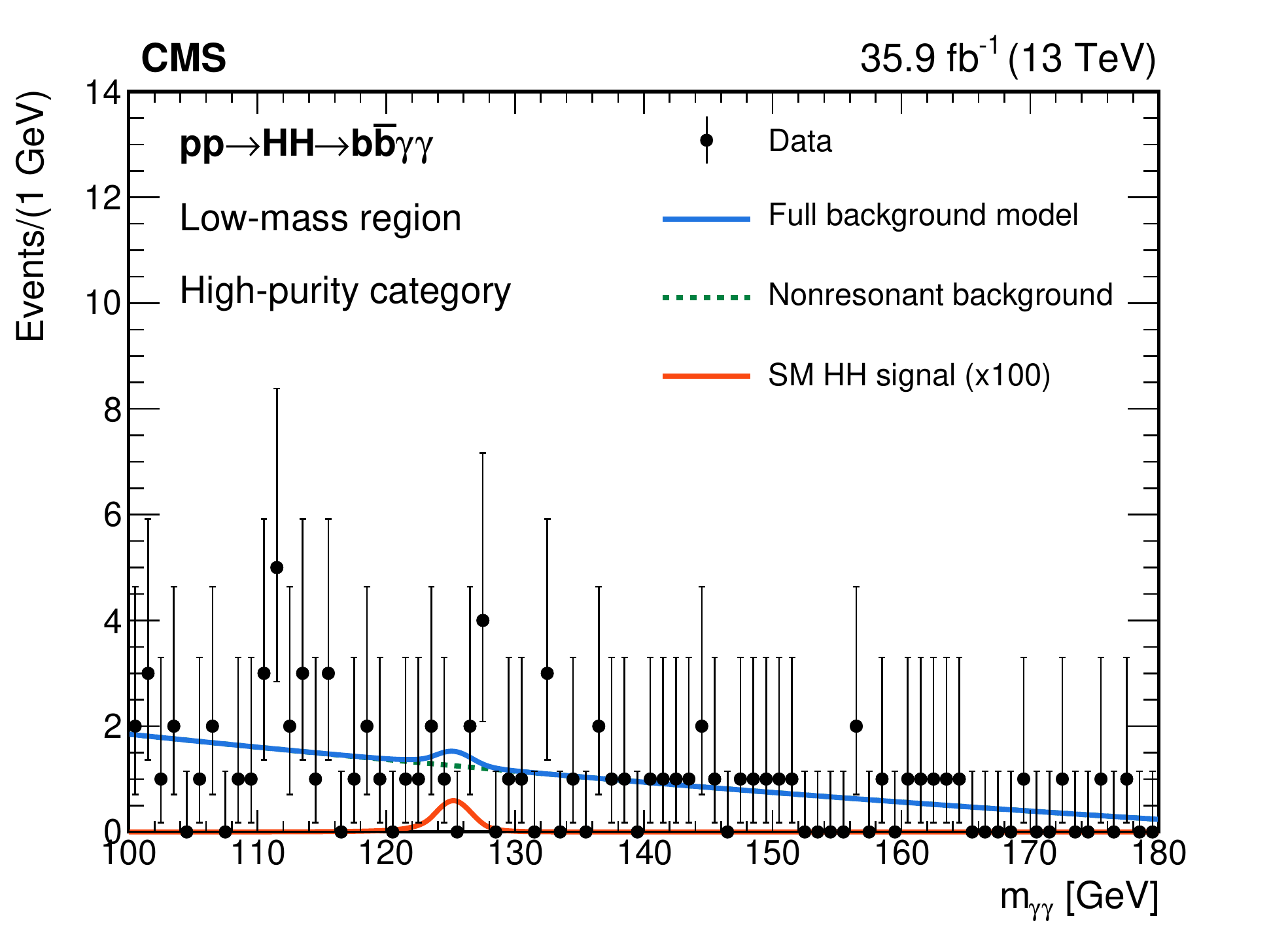}\hfil
  \includegraphics[width=\plotsize\textwidth]{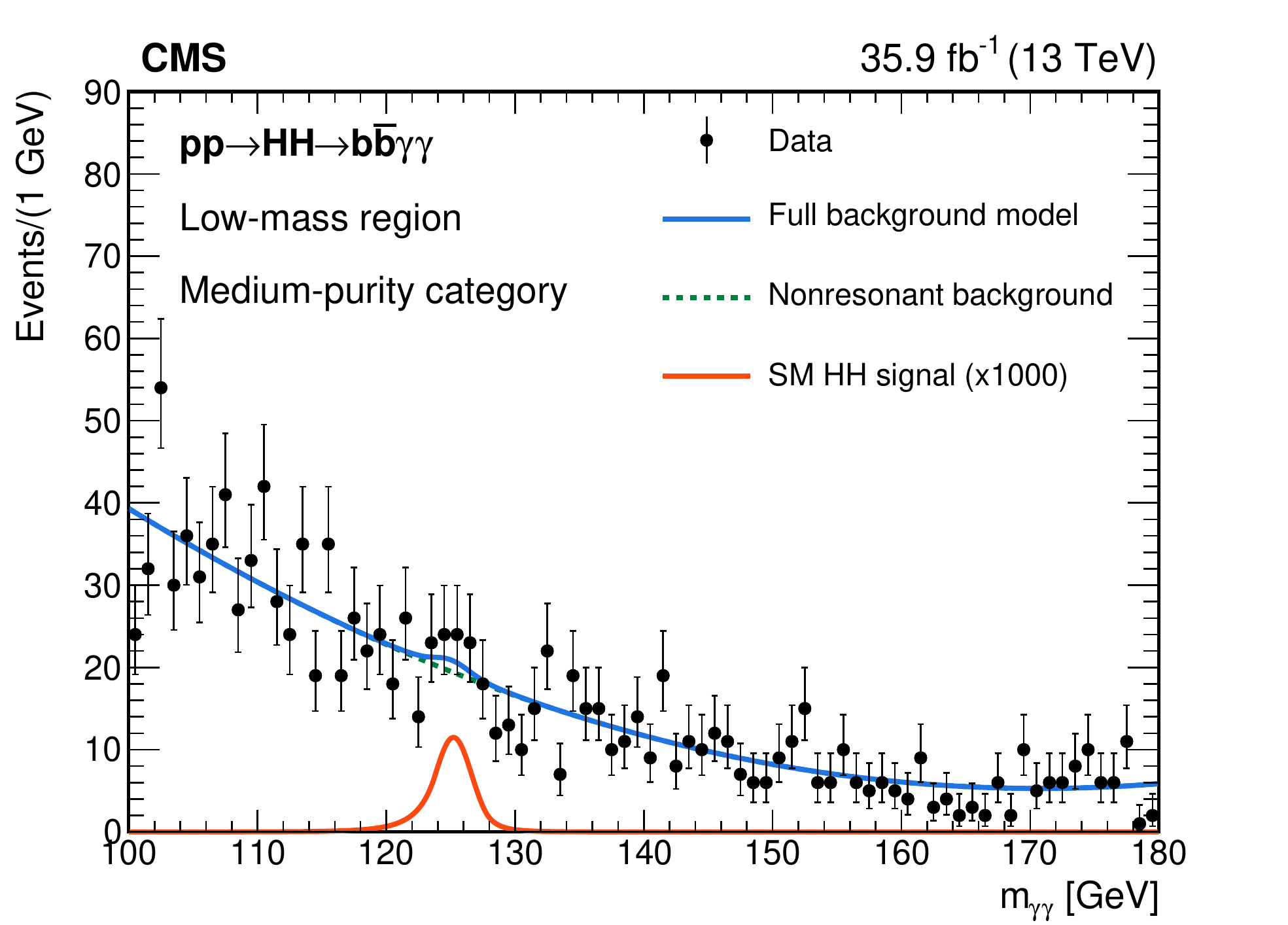}\hfil
  \includegraphics[width=\plotsize\textwidth]{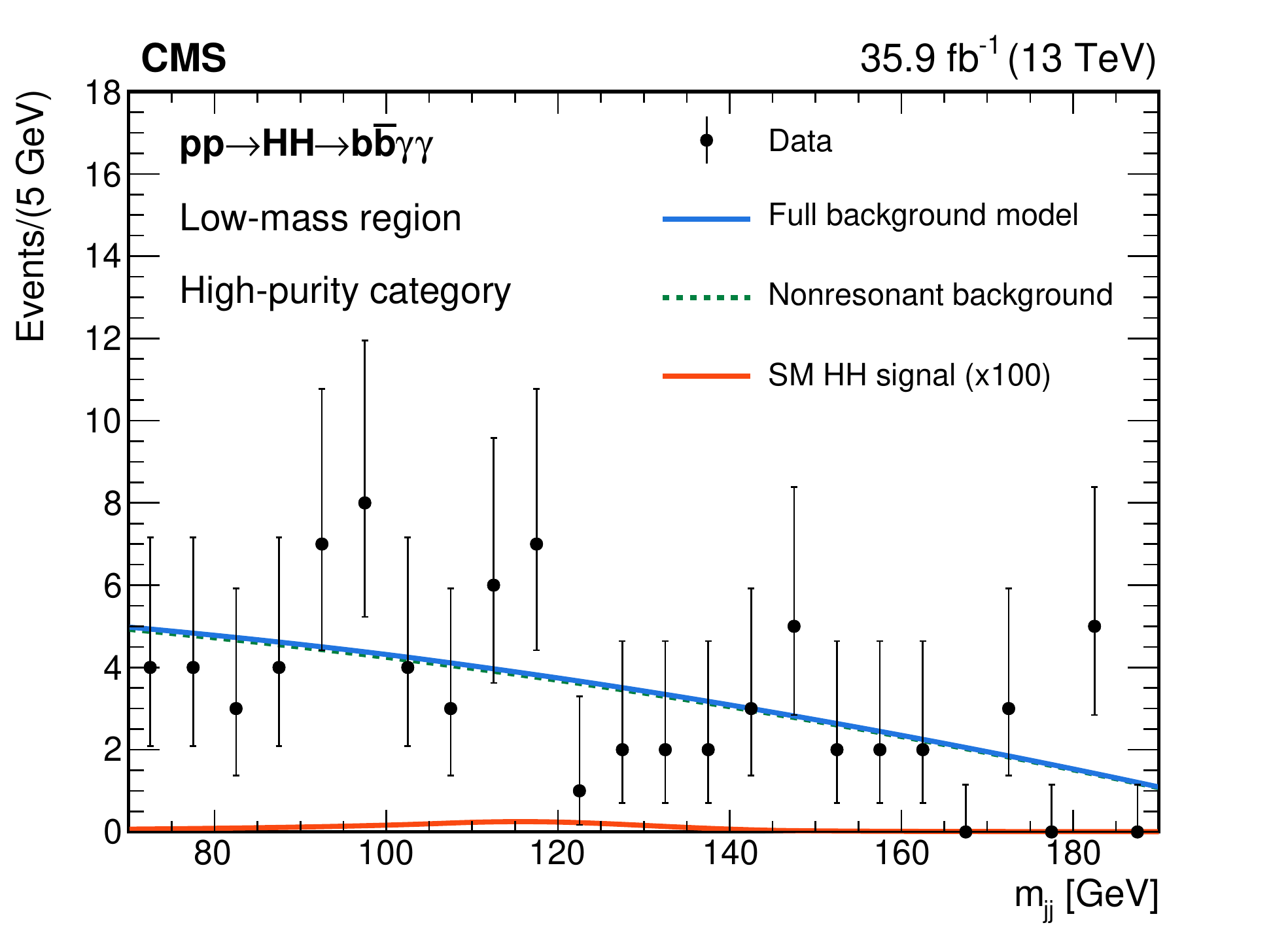}\hfil
  \includegraphics[width=\plotsize\textwidth]{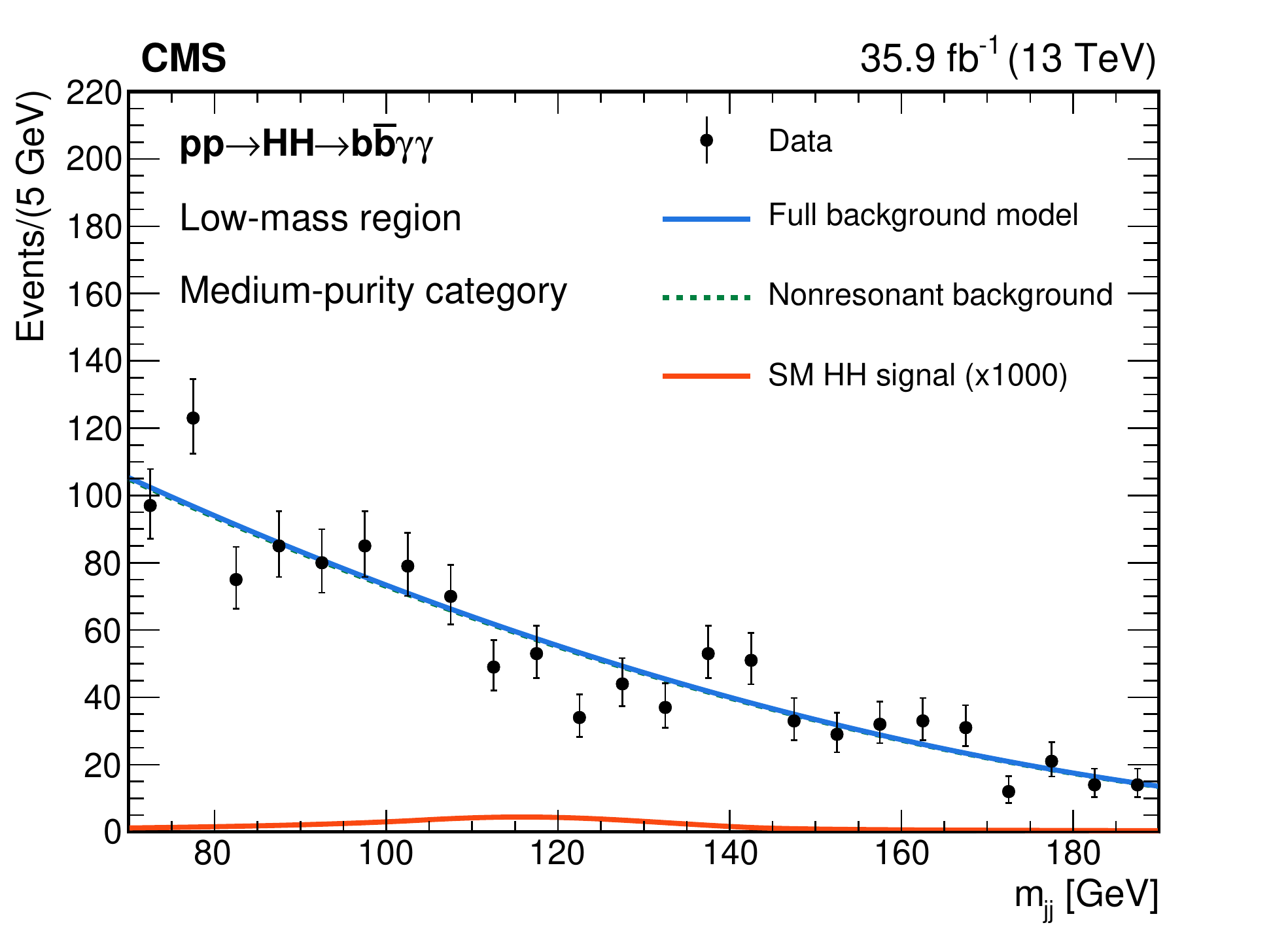}\hfil
  \caption{Background fits for the SM \HH nonresonant analysis selection in the LM region. The plots on the left (right) show the distributions in the HPC (MPC) region. Top plots show the $\Mgg$ spectra and bottom ones the $\Mjj$ ones. The green dashed line represents the nonresonant part of the expected background; the solid blue line represents the full background modeling PD, \ie, the sum of nonresonant background and SM single-Higgs boson contributions scaled to their cross sections;
and the solid red line represents the SM-like \HH production, normalized to its SM cross section times a scaling factor specified in the legend.}
  \label{fig:bkg_fit_nonres_1}
\end{figure*}

The SM single-Higgs boson background contribution is estimated using a PD fitted to simulated samples.
For all production mechanisms, the $\Mgg$ distribution is modeled by a double-sided CB function. The $\Mjj$ modeling depends on the production mechanism: for \ggH and \VBFH production $\Mjj$ is modeled with a Bernstein polynomial; for \VH production a double-sided CB function is expected to describe the line shape of the hadronic decays of vector bosons; for \ttH and \bbH a double-sided CB function is also used.
Like the signal modeling, the final 2D SM single-Higgs boson model is an independent product of models of the $\Mgg$ and $\Mjj$ distributions.
This background contribution is explicitly considered only for the nonresonant search, since for the resonant one it is severely reduced by a tight selection window on $\Mtilde$. The residual events are accounted for by the continuum background models for the $\Mgg$ and $\Mjj$ variables. The one-dimensional projections of the background-plus-signal fits in the signal regions of the nonresonant analysis are shown in Figs. \ref{fig:bkg_fit_nonres_0} and \ref{fig:bkg_fit_nonres_1}.

\section{Fitting procedure and systematic uncertainties}
\label{sec:sys}

A likelihood function is defined based on the total PD including the backgrounds, signal hypothesis, and the data.
Then an unbinned maximum likelihood fit is performed to the 2D \mbox{$\Mgg$\,--\,$\Mjj$} data distribution.
The parameters
for the signal yield and for the background-only PD are constrained in the fit.
Uniform priors are used to parametrize the nonresonant background PD and log-normal priors are assumed for the single-Higgs boson
background parametrizing our degree of uncertainty in the exact SM production cross section.
When converting the
fitted yields into production cross sections, we use simulation to estimate the selection efficiency for the signal.
The difference between the simulation and the data is taken into account
through parameters included in the likelihood function.
Parameters related to the systematic uncertainties (nuisance parameters) are varied in the fit according to a log-normal PD function and can be classified according to their impact on the analysis. The uncertainty in the estimation of the integrated luminosity modifies the total expected signal normalization and is taken to be 2.5\%~\cite{CMS-PAS-LUM-17-001}. Other systematic uncertainties modify the efficiency of the signal selection or impact the signal or the Higgs boson PD.

The photon-related uncertainties are discussed in Ref.~\cite{Khachatryan:2014ira}.
The photon energy scale (PES) is known at
a sub-percent level and the photon energy resolution (PER) is known with 5\% precision.
A 2\% normalization uncertainty is estimated in the offline diphoton selection efficiency
and in the trigger efficiency, while 1\% is assigned to quantify the uncertainty on the photon identification efficiency.

The uncertainties in the jet energy scale (JES) and jet energy resolution (JER) are accounted for by changing the jet response by one standard deviation for each source \cite{JINST6}. They impact the average $\Mjj$ peak position by approximately 1\% and the peak resolution by 5\%. The effects on the signal acceptance are also accounted for.

To use the \PQb tagging score as an input to the classification MVA, its simulated distribution is matched to data by applying differential scale factors that depend on the jet $\pt $ and $\eta$ \cite{BTV-16-002}.
The uncertainty in the efficiency of the categorization MVA is estimated by varying the \PQb tagging differential scale factors within one standard deviation of their uncertainties \cite{BTV-16-002}.
The impacts of PES, PER, JES, and JER on the MVA classification procedure have been found to be negligible.
Those four sources have, nevertheless, a mild impact on the $\Mtilde$--based categorization.

Theoretical uncertainties have been applied to the normalization of the single-Higgs boson background, but not on BSM signals.
When we consider the SM-like search and parametrize the BSM cross section $\sigmaHHBSM$ by the ratio $\muHH = \sigmaHHBSM/\sigmaHHSM$, the theoretical uncertainties on $\sigmaHHSM$ are included in the likelihood. The following sources are considered correlated for all the single-Higgs boson channels (except \bbH) and the double-Higgs boson channel in line with the recommendations from Ref.~\cite{deFlorian:2016spz}: the scale dependence of higher-order terms; the impact from the choice of PDF quadratically summed with the uncertainty on $\alpS$; and, in the case of the \HH channel, the uncertainties related to the inclusion of $\mt$ into the cross section calculations. Finally, for the $\bbH$ channel all sources are summed together including also the uncertainty on the $\PQb$ quark mass.

The systematic uncertainties are summarized in Table~\ref{table:espsyst}. The correlations between different categories are taken into account. The analysis is limited by the statistical precision. For example, the systematic uncertainties worsen the expected cross section limits by 3\% for the search performed assuming a SM-like signal.

\begin{table*}[htb]
\centering
\topcaption{ Summary of systematic uncertainties.  }
\newcolumntype{I}{>{\hspace{5mm}}{l}}
\begin{tabular}{Icc}
\hline
\multicolumn{1}{l}{Sources of systematic uncertainties} & Type & Value (\%)  \\ \hline
\multicolumn{1}{l}{Integrated luminosity} & Normalization & 2.5  \\[\cmsTabSkip]
\multicolumn{3}{l}{Photon related uncertainties} \\
 Diphoton selection (with trigger uncertainties and PES) & Normalization &2.0   \\
 Photon identification & Normalization & 1.0  \\
 PES ($\frac{\Delta \Mgg}{\Mgg}$) & Shape & 0.5  \\
 PER ($\frac{\Delta \sigma_{\gamma\gamma}}{\sigma_{\gamma\gamma}}$) & Shape & 5.0  \\[\cmsTabSkip]
\multicolumn{3}{l}{Jet related uncertainties}  \\
 Dijet selection (JES+JER) & Normalization & 0.5 \\
 JES ($\frac{\Delta \Mjj}{\Mjj}$) & Shape & 1.0  \\
 JER ($\frac{\Delta \sigma_{jj}}{\sigma_{jj}}$) & Shape & 5.0  \\
 Resonant analysis specific uncertainties & & \\
 Mass window selection  (JES+JER) & Normalization & 3.0  \\

 Classification MVA -- \PQb tagging (HPC) & Normalization & 10--19  \\
 Classification MVA -- \PQb tagging (MPC) & Normalization & 3--9  \\[\cmsTabSkip]
\multicolumn{3}{l}{Nonresonant analysis specific uncertainties}  \\
 $\Mtilde$ Classification & Normalization & 0.5  \\
 Classification MVA  -- \PQb tagging (HPC) & Normalization & 10--19  \\
 Classification MVA  -- \PQb tagging (MPC) & Normalization & 3--9 \\[\cmsTabSkip]
\multicolumn{3}{l}{Theoretical uncertainties in the SM single-Higgs boson production} \\
 QCD missing orders (\ggH, \VBFH, \VH, \ttH) & Normalization & 0.4--5.8 \\
 PDF and $\alpS$ uncertainties (\ggH, \VBFH, \VH, \ttH) & Normalization & 1.6--3.6 \\
 Theoretical uncertainty \bbH & Normalization & 20 \\[\cmsTabSkip]
\multicolumn{3}{l}{Theoretical uncertainties in the SM $\HH$ boson production}  \\
 QCD missing orders & Normalization & 4.3--6 \\
 PDF and $\alpS$ uncertainties & Normalization & 3.1 \\
 $\mt$ effects & Normalization & 5 \\
 \hline
\end{tabular}
\label{table:espsyst}
\end{table*}

\section{Results}
\label{section:results}

No evidence for $\HH$ production is observed in the data.
Upper limits on the production cross section of a pair of Higgs bosons times the branching fraction $\mathcal{B}(\HH\to\bbgg)$ are computed using the modified frequentist approach for confidence levels (\CLs), taking the profile likelihood
as a test statistic~\cite{CLS1,CLS2,CLSA,CMS-NOTE-2011-005} in the asymptotic approximation. The limits are subsequently compared
to theoretical predictions assuming SM branching fractions for Higgs boson decays.

\subsection{Resonant signal}

\begin{figure}[hbt!]
\centering
\includegraphics[width=0.49\textwidth]{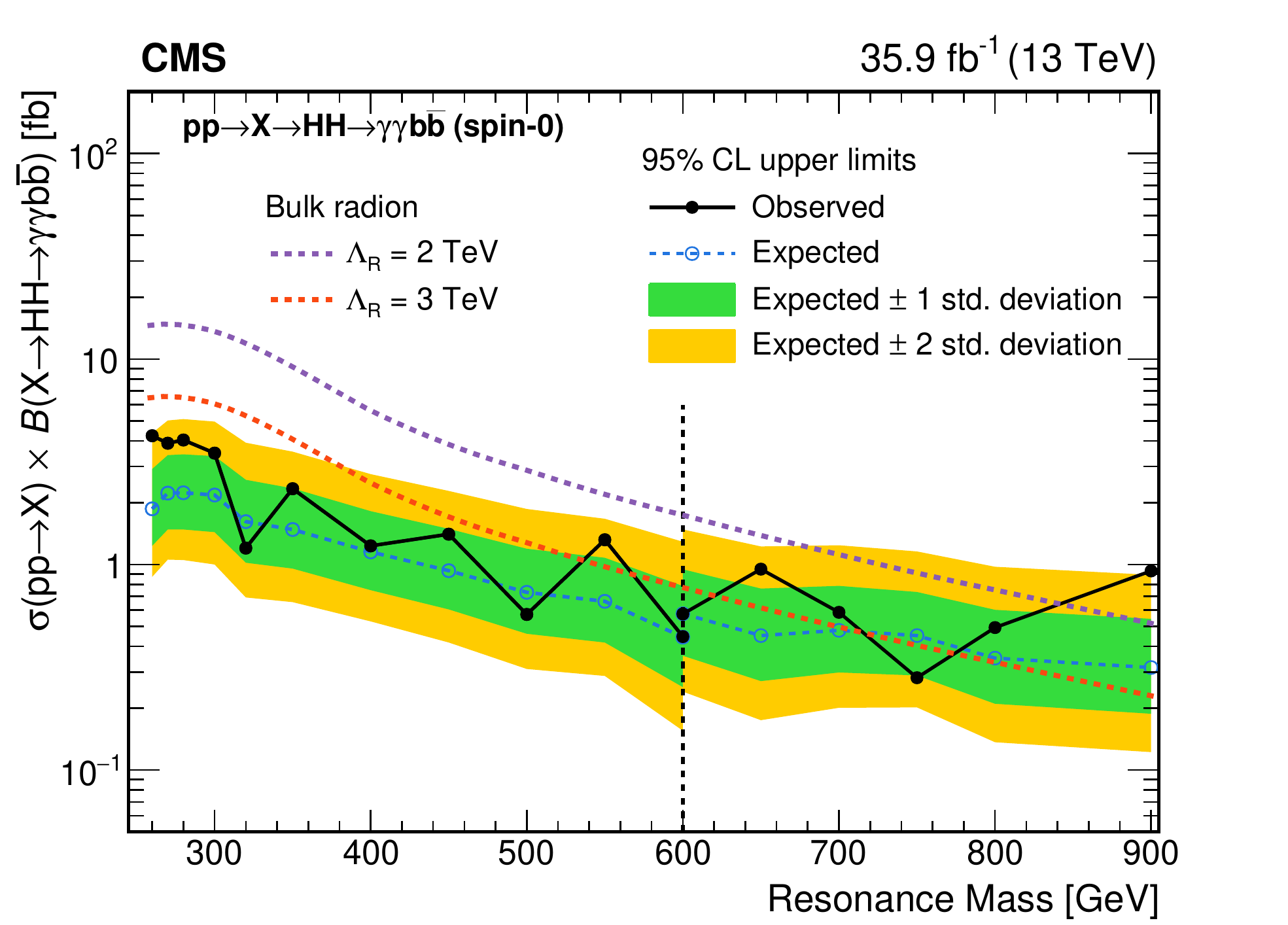}
\includegraphics[width=0.49\textwidth]{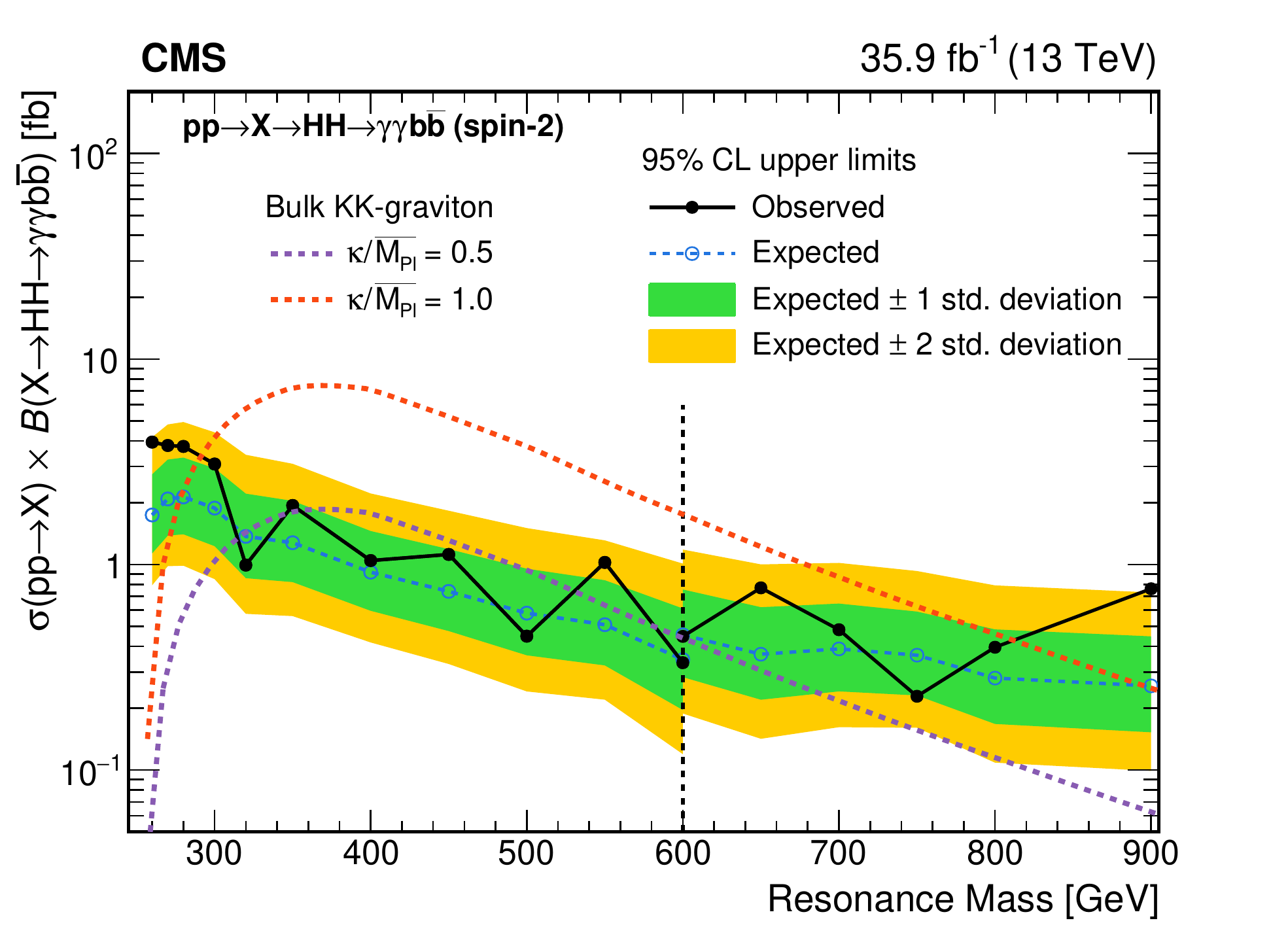}
\caption{
Observed and expected 95\% \CL upper limits on the product of cross section and
  branching fraction $\sigma(\Pp\Pp \to \text{X}) \mathcal{B}(\text{X} \to \HH \to \gamma\gamma \bbbar)$
  obtained through a combination of the two analysis categories (HPC and MPC) for spin-0 (\cmsLeft) and spin-2 (\cmsRight) hypotheses. The green and yellow bands represent, respectively, the
one and two standard deviation extensions beyond the expected
limit.  Also shown are theoretical predictions corresponding to WED models for bulk radions (top) and bulk KK gravitons (bottom).
The vertical dashed lines show the boundary between the low- and high-mass regions. The limits for $\mx = 600\GeV$ are shown for both methods.
}
\label{figure:ExpectedLimits_res}\end{figure}

The observed and median expected upper limits at 95\% confidence level (\CL)
are shown in Fig.~\ref{figure:ExpectedLimits_res}, for the \ppXHHbbgg process assuming spin-0 and a spin-2 resonances.
The data exclude a cross section of 0.23 to 4.2\unit{fb} depending on $\mx$ and the spin hypothesis.

The results are compared with the cross sections for bulk radion and bulk KK graviton production in WED models.
In analogy with the Higgs boson, the hypothesized bulk radion field is expected to be predominantly produced through gluon-gluon fusion~\cite{Mahanta:2000zp} and the cross section is calculated at NLO accuracy in QCD, using the recipe suggested in Ref.~\cite{Giudice:2000av}. The theoretical input used in this letter is identical to the one from a previous CMS analysis \cite{Khachatryan:2016sey}. More details can be found in Ref.~\cite{Oliveira:2014kla}.
The production cross section in this model is proportional to $1/\LambdaR^2$, where $\LambdaR$ is the scale parameter of the theory.
The analysis at $\sqrt{s} = 8\TeV$ \cite{Khachatryan:2016sey} already excluded a radion resonance up to 980\GeV for the scale parameter $\LambdaR = 1\TeV$, but had no sensitivity for $\LambdaR = 3\TeV$.
In this analysis the observed limits are able to exclude radion resonances, assuming $\LambdaR = 2\TeV$, for all points below $\mx = 840\GeV$ and, assuming $\LambdaR = 3\TeV$, below $\mx = 540\GeV$.
The couplings of the gravitons to matter fields are defined by $\koM$, with $\overline{\Mpl} \equiv \Mpl /\sqrt{8\pi}$ being the reduced Planck mass and $\kappa$ the warp factor of the metric.
Assuming $\koM = 1.0$, gravitons are excluded within the range $290 < \mx < 810\GeV$, while assuming $\koM = 0.5$, the exclusion range is $350 < \mx < 530\GeV$.

\begin{figure}[hbt!]
\centering
\includegraphics[width=\cmsFigWidth]{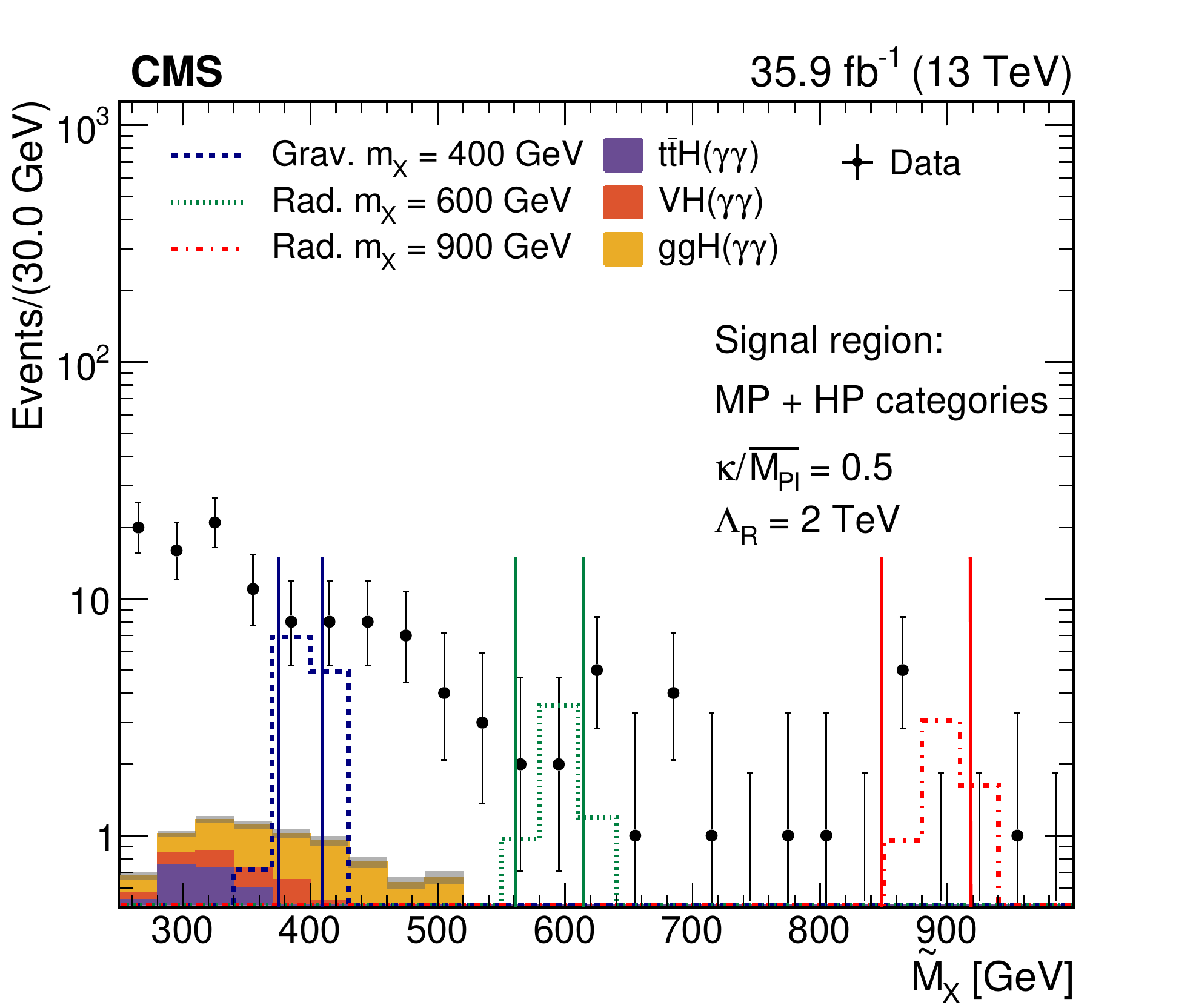}
\caption{ Data (dots) with statistical uncertainties (vertical black lines around the dots) histogrammed in bins of $\Mtilde$ are overlaid onto expected Monte Carlo simulated distributions for three single-Higgs bosons samples (\ttH, \VH, and \ggH) and three different resonant signal hypotheses. The data-driven backgrounds, dominated by $\QCD$, are not shown and are specific to the analysis categories. The events are selected in the signal region defined by $122 < \Mgg < 128\GeV$ and $90 < \Mjj < 160\GeV$ and use the photons and jets selections summarized in Table \ref{table:gencut}. The MP and HP categories are merged together. The signals are normalized to the theory cross section calculated with the parameters shown in the figure and assuming a narrow width approximation. The vertical lines shows the mass windows that are used to select in $\mx$ and set the limits shown in Fig. \ref{figure:ExpectedLimits_res}
}
\label{figure:resonant}\end{figure}

The events in a signal region defined by $122 < \Mgg < 128\GeV$ and $90 < \Mjj < 160\GeV$ are shown as a function of $\mx$ in Fig. \ref{figure:resonant} and compared to three different resonant signal hypotheses with theory parameters chosen in such a way that their cross sections are close to the excluded ones. The vertical lines show the mass windows that are used to select in $\mx$ and set the limits shown in Fig. \ref{figure:ExpectedLimits_res}.

\subsection{Nonresonant signal}

The observed (expected) 95\% \CL upper limit on the SM-like
$\ppHHbbgg$ process is 2.0 (1.6\unit{fb}), and 0.79 (0.63\unit{pb}) for the total  $\ggHH$ production cross section assuming SM Higgs boson branching fractions.
The results can also be interpreted in terms of observed (expected) upper limits on $\muHH$ of 24 (19). The constraint on $\muHH$ is a factor of three more restrictive that the previous search \cite{Khachatryan:2016sey}.

An additional study is performed including both VBF \HH and \ggHH production mechanisms in the definition of the scaling factor
\begin{equation}
\muHHext = \frac{\sigmaHHBSM +\sigmaHHBSMVBF}{\sigmaHHSM+\sigmaHHSMVBF}
\end{equation}
where $\sigmaHHSMVBF = 1.64^{+0.05}_{-0.06}\unit{fb}$ \cite{deFlorian:2016spz}.
The expected sensitivity of the analysis for $\muHHext$ improves by 1.3\% compared to $\muHH$. The improvement is smaller than the relative contribution of the VBF production cross section to the total one in the SM because of the nonoptimal selection efficiency of this analysis for the VBF events, as explained in Section \ref{sec:AnalysisMethods}.

\begin{figure}[!htb]
  \centering
\includegraphics[width=\cmsFigWidth]{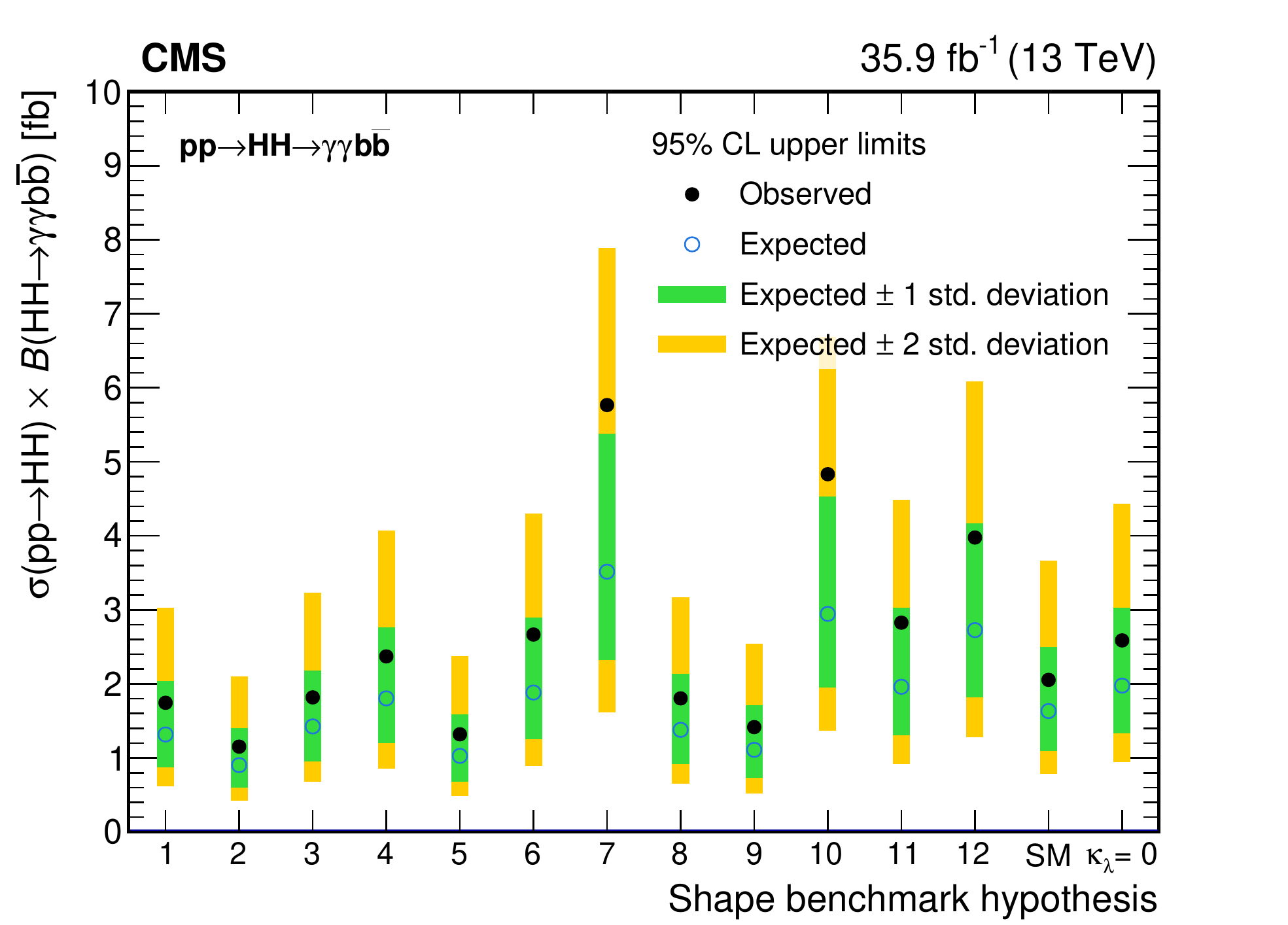}
\includegraphics[width=\cmsFigWidth]{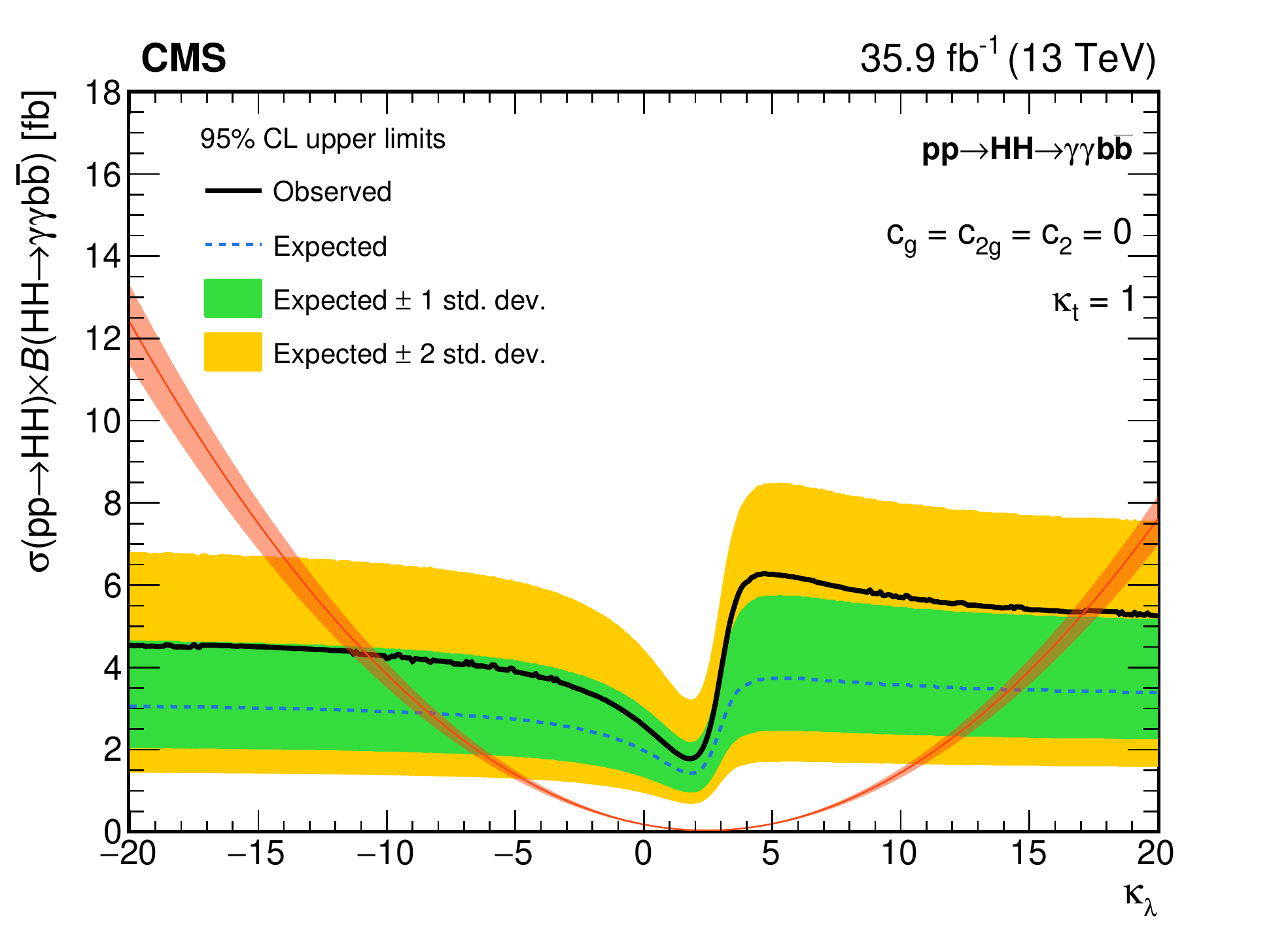}

  \caption{Expected and observed 95\% \CL upper limits on the SM-like \HH production cross section times $\mathcal{B}(\HH\to\bbgg)$ obtained for different nonresonant benchmark models (defined in Table \ref{tab:bench}) (top); for different values of the $\kapl$ (bottom). The green and yellow bands represent, respectively, the one and two standard deviation extensions beyond the expected limit.
The red line in the bottom plot shows the prediction of theory with the associated uncertainties shown as the orange band.
}
  \label{fig:NRlim_lambda}
\end{figure}

The results are also interpreted in the context of Higgs boson anomalous couplings.
Limits on the different BSM benchmark hypotheses (listed in Table \ref{tab:bench}) are shown in Fig. \ref{fig:NRlim_lambda} (top).
Using the limits on the benchmark hypotheses and the map between the clusters and the points in the 5D BSM parameter space, one can estimate the constraints provided by these data in different regions of phase space.
It is important to stress that the same analysis categories are used for the SM-like search and for all BSM nonresonant searches. The differences between the limits come only from the kinematic properties of the benchmark signal hypotheses. For instance, the tightest constraint is placed on the benchmark 2 hypothesis, which features a $\mHH$ spectrum that extends above 1\TeV owing to large contributions from dimension-6 operators. The least restrictive constraint is on benchmark 7 that describes models with large values of $\kapl$, where the $\mHH$ spectrum peaks below 300\GeV. For benchmark 2 most of the events would be observed in the HM categories, while for benchmark 7 most events fall in the LM categories.
Assuming equal cross sections, benchmark hypothesis 2 has a much better signal-over-background ratio than benchmark hypothesis 7.
In the intermediate case of SM-like production it was observed that the HM categories have the best sensitivity.

We reweight the benchmark samples to model different values of the SM coupling modifier $\kapl$, with $\kapt$ and other BSM parameters fixed to their SM values. In Fig.~\ref{fig:NRlim_lambda} (bottom),
95\% \CL limits on the nonresonant Higgs boson pair production cross sections are shown as a function of $\kapl$.
Assuming that the top quark Yukawa coupling is SM-like ($\kapt = 1$), the analysis constrains $\kapl$ to values between $-11$ and $17$.

\section{Summary}

A search is performed by the CMS Collaboration for the resonant and nonresonant
production of a pair of Higgs bosons in the decay channel $\HH \to \bbgg$, based on an integrated
luminosity of 35.9\fbinv of $\Pp\Pp$ collisions
collected at $\sqrt{s}=13\TeV$ in 2016.
No statistically significant deviations from the standard model (SM) predictions are found.
Upper limits at a 95\% \CL are set on the cross
sections for the production of new particles decaying to two Higgs bosons in the mass range between 250 and 900\GeV, under the spin-0 and spin-2 hypotheses.
In the case of beyond SM predictions, based
on the assumption of the existence of a warped extra dimension, we exclude the radion (spin-0) signal hypothesis, assuming the scale parameter $\LambdaR = 3\TeV$, for all masses below $\mx = 540\GeV$, and the KK graviton (spin-2) hypothesis for the mass range $290 < \mx < 810\GeV$, assuming $\koM = 1.0$ ($\overline{\Mpl}$ being the reduced Planck mass and $\kappa$ the warp factor of the metric).
For nonresonant production with SM-like kinematics, a 95\% \CL
upper limit of 2.0\unit{fb} is set on $\sigma(\Pp\Pp\to\HH \to\bbgg)$, corresponding to about 24 times the SM prediction.
Anomalous couplings of the Higgs boson are also investigated, as well as the vector boson fusion $\HH$ production process.
Values of the effective Higgs boson self-coupling $\kapl$ are constrained to be within the range $-11 < \kapl < 17$ at 95\% \CL, assuming all other Higgs boson couplings to be at their SM values.
The direct constraints reported on $\kapl$ are the most restrictive to date.

\begin{acknowledgments}

We congratulate our colleagues in the CERN accelerator departments for the excellent performance of the LHC and thank the technical and administrative staffs at CERN and at other CMS institutes for their contributions to the success of the CMS effort. In addition, we gratefully acknowledge the computing centers and personnel of the Worldwide LHC Computing Grid for delivering so effectively the computing infrastructure essential to our analyses. Finally, we acknowledge the enduring support for the construction and operation of the LHC and the CMS detector provided by the following funding agencies: BMWFW and FWF (Austria); FNRS and FWO (Belgium); CNPq, CAPES, FAPERJ, and FAPESP (Brazil); MES (Bulgaria); CERN; CAS, MoST, and NSFC (China); COLCIENCIAS (Colombia); MSES and CSF (Croatia); RPF (Cyprus); SENESCYT (Ecuador); MoER, ERC IUT, and ERDF (Estonia); Academy of Finland, MEC, and HIP (Finland); CEA and CNRS/IN2P3 (France); BMBF, DFG, and HGF (Germany); GSRT (Greece); OTKA and NIH (Hungary); DAE and DST (India); IPM (Iran); SFI (Ireland); INFN (Italy); MSIP and NRF (Republic of Korea); LAS (Lithuania); MOE and UM (Malaysia); BUAP, CINVESTAV, CONACYT, LNS, SEP, and UASLP-FAI (Mexico); MBIE (New Zealand); PAEC (Pakistan); MSHE and NSC (Poland); FCT (Portugal); JINR (Dubna); MON, RosAtom, RAS, RFBR and RAEP (Russia); MESTD (Serbia); SEIDI, CPAN, PCTI and FEDER (Spain); Swiss Funding Agencies (Switzerland); MST (Taipei); ThEPCenter, IPST, STAR, and NSTDA (Thailand); TUBITAK and TAEK (Turkey); NASU and SFFR (Ukraine); STFC (United Kingdom); DOE and NSF (USA).

\hyphenation{Rachada-pisek} Individuals have received support from the Marie-Curie program and the European Research Council and Horizon 2020 Grant, contract No. 675440 (European Union); the Leventis Foundation; the A. P. Sloan Foundation; the Alexander von Humboldt Foundation; the Belgian Federal Science Policy Office; the Fonds pour la Formation \`a la Recherche dans l'Industrie et dans l'Agriculture (FRIA-Belgium); the Agentschap voor Innovatie door Wetenschap en Technologie (IWT-Belgium); the F.R.S.-FNRS and FWO (Belgium) under the "Excellence of Science - EOS" - be.h project n. 30820817; the Ministry of Education, Youth and Sports (MEYS) of the Czech Republic; the Council of Science and Industrial Research, India; the HOMING PLUS program of the Foundation for Polish Science, cofinanced from European Union, Regional Development Fund, the Mobility Plus program of the Ministry of Science and Higher Education, the National Science Center (Poland), contracts Harmonia 2014/14/M/ST2/00428, Opus 2014/13/B/ST2/02543, 2014/15/B/ST2/03998, and 2015/19/B/ST2/02861, Sonata-bis 2012/07/E/ST2/01406; the National Priorities Research Program by Qatar National Research Fund; the Programa Severo Ochoa del Principado de Asturias; the Thalis and Aristeia programs cofinanced by EU-ESF and the Greek NSRF; the Rachadapisek Sompot Fund for Postdoctoral Fellowship, Chulalongkorn University and the Chulalongkorn Academic into Its 2nd Century Project Advancement Project (Thailand); the Welch Foundation, contract C-1845; and the Weston Havens Foundation (USA).

\end{acknowledgments}

\bibliography{auto_generated}

\cleardoublepage \appendix\section{The CMS Collaboration \label{app:collab}}\begin{sloppypar}\hyphenpenalty=5000\widowpenalty=500\clubpenalty=5000\vskip\cmsinstskip
\textbf{Yerevan~Physics~Institute,~Yerevan,~Armenia}\\*[0pt]
A.M.~Sirunyan, A.~Tumasyan
\vskip\cmsinstskip
\textbf{Institut~f\"{u}r~Hochenergiephysik,~Wien,~Austria}\\*[0pt]
W.~Adam, F.~Ambrogi, E.~Asilar, T.~Bergauer, J.~Brandstetter, E.~Brondolin, M.~Dragicevic, J.~Er\"{o}, A.~Escalante~Del~Valle, M.~Flechl, M.~Friedl, R.~Fr\"{u}hwirth\cmsAuthorMark{1}, V.M.~Ghete, J.~Grossmann, J.~Hrubec, M.~Jeitler\cmsAuthorMark{1}, A.~K\"{o}nig, N.~Krammer, I.~Kr\"{a}tschmer, D.~Liko, T.~Madlener, I.~Mikulec, E.~Pree, N.~Rad, H.~Rohringer, J.~Schieck\cmsAuthorMark{1}, R.~Sch\"{o}fbeck, M.~Spanring, D.~Spitzbart, A.~Taurok, W.~Waltenberger, J.~Wittmann, C.-E.~Wulz\cmsAuthorMark{1}, M.~Zarucki
% ...
\vskip\cmsinstskip
% ...
1:~Also at~Vienna~University~of~Technology,~Vienna,~Austria\\
% ...
\end{sloppypar}
\end{document}